\documentclass{aastex}
\usepackage{emulateapj5}
\usepackage{timesfonts}

\makeatletter

\newenvironment{inlinefigure}{%
\def\@captype{figure}%
\noindent\begin{minipage}{0.999\linewidth}\begin{center}}
{\end{center}\end{minipage}\smallskip}
\makeatother

\begin{document}   

\lefthead{NAGASHIMA ET AL.}
\righthead{GALAXY NUMBER COUNTS IN THE SUBARU DEEP FIELD}

\slugcomment{To appear in ApJ}

\title{Galaxy Number Counts in the Subaru Deep Field: Multi-band
Analysis in a Hierarchical Galaxy Formation Model}

\author{Masahiro Nagashima,\altaffilmark{1}
 Yuzuru Yoshii,\altaffilmark{2}$^{,}$\altaffilmark{3}
 Tomonori Totani,\altaffilmark{1,4} and
 Naoteru Gouda\altaffilmark{1}}
\email{masa@th.nao.ac.jp}
\altaffiltext{1}{National Astronomical Observatory, Mitaka, Tokyo 181-8588, 
Japan}
\altaffiltext{2}{Institute of Astronomy, Graduate School of Science, The 
University 
of Tokyo, 2-21-1 Osawa, Mitaka, Tokyo 181-8588, Japan}
\altaffiltext{3}{Research Center for the Early Universe, Graduate School of 
Science, The University of Tokyo, 7-3-1 Hongo, Bunkyo-ku, Tokyo 113-0033, 
Japan}
\altaffiltext{4}{Peyton Hall, Princeton University, Princeton, NJ08544-1001, USA}

\begin{abstract}   
Number counts of galaxies are re-analyzed using a semi-analytic model
(SAM) of galaxy formation based on the hierarchical clustering scenario.
Faint galaxies in the Subaru Deep Field (SDF, near-infrared $J$ and
$K'$) and the Hubble Deep Field (HDF, ultraviolet/optical $U$, $B$, $V$,
and $I$) are compared with our model galaxies.  We have determined the
astrophysical parameters in the SAM that reproduce observations of
nearby galaxies, and used them to predict the number counts and
redshifts of faint galaxies for three cosmological models, the standard
cold dark matter (CDM) universe, a low-density flat universe with
nonzero cosmological constant, and a low-density open universe with zero
cosmological constant.  The novelty of our SAM analysis is the inclusion
of selection effects arising from the cosmological dimming of surface
brightness of high-redshift galaxies, and from the absorption of visible
light by internal dust and intergalactic \ion{H}{1} clouds.  As was
found in our previous work, in which the ultraviolet/optical HDF
galaxies were compared with our model galaxies, we find that our SAM
reproduces counts of near-infrared SDF galaxies in a low-density
universe either with or without a cosmological constant, and that the
standard CDM universe is {\it not} preferred, as suggested by other
recent studies.  Moreover, we find that simple prescriptions for (1) the
timescale of star formation being proportional to the dynamical time
scale of the formation of galactic disks, (2) the size of galactic disks
being rotationally supported with the same specific angular momentum as
that of surrounding dark halo, and (3) the dust optical depth being
proportional to the metallicity of cold gas, cannot completely explain
all of observed data.  Improved prescriptions incorporating mild
redshift-dependence for those are suggested from our SAM analysis.
\end{abstract}   
   
\keywords{cosmology: theory -- galaxies: evolution -- galaxies:
  formation -- large-scale structure of universe }

\section{INTRODUCTION}   

It is well known that the number of faint galaxies in a given area of
sky can constrain cosmological parameters, because it depends on the
geometry of the Universe (e.g., Peebles 1993).  Many efforts have been
devoted to this subject using traditional galaxy evolution models
assuming monolithic collapse, such as the wind model for elliptical
galaxies and the infall model for spiral galaxies (e.g., Yoshii \&
Takahara 1988).  These models are in fact able to reproduce many of the
observed properties of nearby galaxies, and provide a useful theoretical
tool for understanding their evolution \citep{ay86, ay87, ayt91}.

In the analyses of galaxy counts from traditional approaches, it has
been found that the Einstein-de Sitter (EdS) universe, a representation
of the standard cold dark matter (CDM) universe, is not reconcilable
with the observed high counts of faint galaxies, and that a low-density
universe is preferred \citep{yt88, yp91, yp95, y93}.  Recently,
\citet{ty00} and \citet{t01} compared their predictions against the
observed number counts to the faint limits in the Hubble Deep Field
(HDF; Williams et al. 1996) and in the Subaru Deep Field (SDF; Maihara
et al. 2001), taking into account various selection effects and allowing
for the possibility of number evolution of galaxies in a
phenomenological way.  Note that the SDF counts are now the deepest
near-infrared ones with the 5$\sigma$ limiting magnitude of $K=23.5$ in
total magnitude.  They confirmed that the EdS universe cannot reproduce
the observed high counts.  However, in their best-fit models, the merger
rates of HDF and SDF galaxies are a little different.  A mild merger
rate is needed to reproduce the counts in the HDF, while a negligible
rate was necessary for the SDF.  It should be noted that the photometric
passbands for the two applications are different; ultraviolet/optical
for the HDF and near-infrared for the SDF.  They suggested that the
difference of the merger rate might be originated by
morphology-dependent number evolution because late-type galaxies are
mainly seen in shorter wavelength such as $B$-band and early-type
galaxies are seen in longer wavelength such as $K$-band.  In any case,
it should be explained why the merger rate depends on the observed
wavelength in order to obtain a better understanding of the galaxy
formation process.

On the other hand, in the study of formation of large-scale structure in
the universe, both theory and observation suggest that gravitationally
bound objects, such as galaxy clusters, are formed through continuous
mergers of dark halos with an initial density fluctuation spectrum
predicted by the CDM models.  Based on this scenario of {\it
hierarchical clustering}, the so-called semi-analytic models (SAMs) of
galaxy formation have been developed by a number of authors (Kauffmann,
White \& Guiderdoni 1993; Cole et al. 1994, 2000; Somerville \& Primack
1999; Nagashima et al. 2001, hereafter NTGY).  SAMs successfully
reproduced a variety of observed features of local galaxies, such as
their luminosity function, color distribution, and so on.

Faint galaxy number counts have also been analyzed using SAMs
\citep{c94, kgw94, h95, bcf96}.  These studies showed that predicted
number counts in the EdS universe agree with the observed counts.  Their
results, however, contradict analyses carried out with traditional
galaxy evolution models.  Recently, this contradiction was resolved by
NTGY, in which their SAM is compared with the galaxy counts in the HDF.
They found, by matching properties of model galaxies with observation
especially in local luminosity functions and cold gas mass fraction,
that normalization of model parameters, related to combinations of
physical processes such as star formation (SF) and supernova (SN)
feedback, is very important, and that accounting for selection effects
caused by cosmological dimming of surface brightness and absorption of
emitted light by internal dust and intergalactic \ion{H}{1} clouds is
crucial in the analysis of galaxy counts, as shown by \citet{ty00}.  It
should be noted that recent analysis by \citet{lypcf} also clarify the
importance of the selection effects caused by the cosmological dimming
of surface brightness in the observational point of view.  They
introduced the star formation rate intensity distribution function,
which was derived from the ultraviolet luminosity density for the HDF
galaxies, at several redshifts and found that at high redshift
significant fraction of ultraviolet luminosity density must be missed
due to the cosmological dimming of surface brightness.

The purpose of this paper is to examine whether our SAM can {\it
simultaneously} reproduce both the ultraviolet/optical and near-infrared
galaxy counts in the HDF and in the SDF.  Because the luminosity of
galaxies in different passbands reflects the influence of different
stellar populations, the subject of multi-band number counts provides a
strong constraint on galaxy formation.  In this paper, using selection
criteria for SDF galaxies based on \citet{t01}, we compare our SAM
prediction of galaxy counts with the observed counts in the SDF.

This paper is outlined as follows. In \S2 we briefly describe our SAM,
which is almost the same as our previous models (NTGY). In \S3 we
constrain the astrophysical parameters in our SAM analysis using local
observations. In \S4 we compare theoretical number counts of faint
galaxies with the HDF and SDF data, and in \S5 we discuss the range of
uncertainties in our calculations of galaxy number counts.  In \S6 we
provide a summary and discussion.

\section{MODEL}\label{sec:model}

The galaxy formation scenario that we model is as follows.  In the CDM
universe, dark matter halos cluster gravitationally, and merge in a
manner that depends on the adopted power spectrum of the initial density
fluctuations.  In each of the merged dark halos, radiative gas cooling,
star formation, and gas reheating by supernovae occur.  The cooled dense
gas and stars constitute {\it galaxies}.  These galaxies sometimes merge
together in a common dark halo, and more massive galaxies form.
Repeating these processes, galaxies form and evolve to the present
epoch.

The SAM analysis we perform obtains essentially the same results of our
previous SAM analyses, with minor differences in a number of details.
In this section we briefly describe our model.  Details of the model we
employ are described in NTGY, except for a few differences, which are
explicitly mentioned below.

\subsection{Scheme of Galaxy Formation}\label{sec:gf}

The merging histories of dark halos are realized by a Monte Carlo method
proposed by \citet{sk99}, based on the extended Press-Schechter
formalism \citep{bcek91, b91, lc93}.  This formalism is an extension of
the Press-Schechter formalism \citep{ps74}, which gives us the mass
function of dark halos, $n(M)dM$, to estimate the mass function of
progenitor halos with mass $M_{1}$ at a redshift $z_{0}+\Delta z$ of a
single dark halo with mass $M_{0}$ collapsing at a redshift $z_{0}$,
$n(M_{1}; z_{0}+\Delta z|M_{0}; z_{0})dM_{1}$.  Dark halos with circular
velocity $V_{\rm circ}\geq 40$km~s$^{-1}$ are regarded as isolated
halos, and otherwise regarded as diffuse accreted matter.  The mean
density in dark halos is assumed to be proportional to the cosmic mean
density at the collapsing epoch using a spherically symmetric collapse
model \citep{t69, gg72}.  Each collapsing dark halo contains baryonic
matter with a mass fraction $\Omega_{\rm b}/\Omega_{0}$, where
$\Omega_{0}$ and $\Omega_{\rm b}$ are the parameters of total and baryon
mass densities relative to the critical cosmic mass density.  We adopt a
value of $\Omega_{\rm b}=0.015h^{-2}$ given by \citet{syb00} in which
they estimated the value from the primordial lithum abundance, where $h$
is the Hubble parameter, $h\equiv H_{0}/100$km~s$^{-1}$Mpc$^{-1}$.  Note
that a recent measurement of the anisotropy of the cosmic microwave
background by the BOOMERANG project suggests a slight higer value,
$\Omega_{\rm b}\simeq 0.02h^{-2}$ \citep{boomerang}.  The effect of
changing $\Omega_{\rm b}$ has already been investigated by \citet{c00}
and they showed that this mainly affects the value of the invisible
stellar mass fraction such as brown dwarfs parametrized by $\Upsilon$
(see the next subsection).  We also checked whether our results are
changed or not in the case of $\Omega_{\rm b}\sim 0.04$ and found that
this does not affect them.  The baryonic matter consists of diffuse hot
gas, dense cold gas, and stars.

When a halo collapses, the hot gas is shock-heated to the virial
temperature of the halo with an isothermal density profile.  A part of
the hot gas cools and accretes to disk of a galaxy until the subsequent
collapse of the dark halos.  The amount of the cold gas involved is
calculated by using metallicity-dependent cooling functions provided by
\citet{sd93}.  The difference of cooling rates between the primordial
and metal-polluted gases is prominent at $T\sim 10^{6}$K due to
line-cooling of metals.  The cooling is, however, very efficient in dark
halos with a virial temperature of $T\sim 10^{6}$K even in the case of
the primordial gas, so the metallicity dependence of cooling rate only
slightly affects our results.  In order to avoid the formation of
unphysically large galaxies, the cooling process is applied only to
halos with $V_{\rm circ}<$500 km~s$^{-1}$ in the standard CDM and 400
km~s$^{-1}$ in low-density universes according to some previous SAMs.
This handling would be needed because the simple isothermal distribution
forms so-called ``monster galaxies'' due to the too efficient cooling at
the center of halos.  While this problem will probably solved by
adopting another isothermal distribution with central core \citep{c00},
we take the above simple approach because this does not cause any
serious problems in estimating galaxy counts.

Stars in the disk are formed from the cold gas.  The SF rate (SFR)
$\dot{M}_{*}$ is given by the cold gas mass $M_{\rm cold}$ and a SF
timescale $\tau_{*}$ as $\dot{M}_{*}=M_{\rm cold}/\tau_{*}$.  We use two
SF models.  One model adopts constant star formation (CSF), in which
$\tau_{*}$ is a constant against redshift; the other adopts dynamical
star formation (DSF), in which $\tau_{*}$ is proportional to the
dynamical timescale of the halo, which allows for the possibility that
the SF efficiency is variable with redshift.  We then express the SF
timescale as
\begin{eqnarray}
\tau_{*}=\left\{
\begin{array}{ll}
\displaystyle{\tau_{*}^{0}\left(\frac{V_{\rm circ}}{300{{\rm 
km~s}^{-1}}}\right)
^{\alpha_{*}}} & \mbox{(CSF)},\\
\displaystyle{\tau_{*}^{0}\left(\frac{V_{\rm circ}}{300{{\rm 
km~s}^{-1}}}\right)
^{\alpha_{*}}\left[\frac{\tau_{\rm dyn}(z)}{\tau_{\rm dyn}(0)}\right]
(1+z)^{\sigma}} & \mbox{(DSF)},
\end{array}
\right.
\end{eqnarray}
where $\tau_{*}^{0}, \alpha_{*}$ and $\sigma$ are free parameters.  The
former two parameters are chosen so as to match the fraction of cold gas
mass with the observed fraction (see next section).  While, of course,
some fraction of cold gas might not be seen, we adopt the gas mass
fraction to constrain the SF-related parameters according to
\citet{c00}.  If much of this cold gas would be hidden, $\tau_{*}^{0}$
should be longer than our value.  Note that in our previous paper (NTGY)
we used DSF only in the case of $\sigma=0$.  In Figure \ref{fig:sfr} we
show the timescales for CSF (dot-dashed line) and DSF with $\sigma=0$
(solid line) and $\sigma=1$ (dashed line) in a $\Lambda$-dominated flat
universe.  Because objects which collapsed at higher redshift have
higher density, the dynamical timescale $\tau_{\rm dyn}$ is shorter at
higher redshift.  Thus, the DSF scenario with smaller $\sigma$ converts
the cold gas into stars more rapidly than the CSF.  In this paper we
examine not only the CSF and $\sigma=0$ DSF models but also the DSF
model with $\sigma=1$ as an intermediate case.

\begin{inlinefigure}
\includegraphics[width=\hsize]{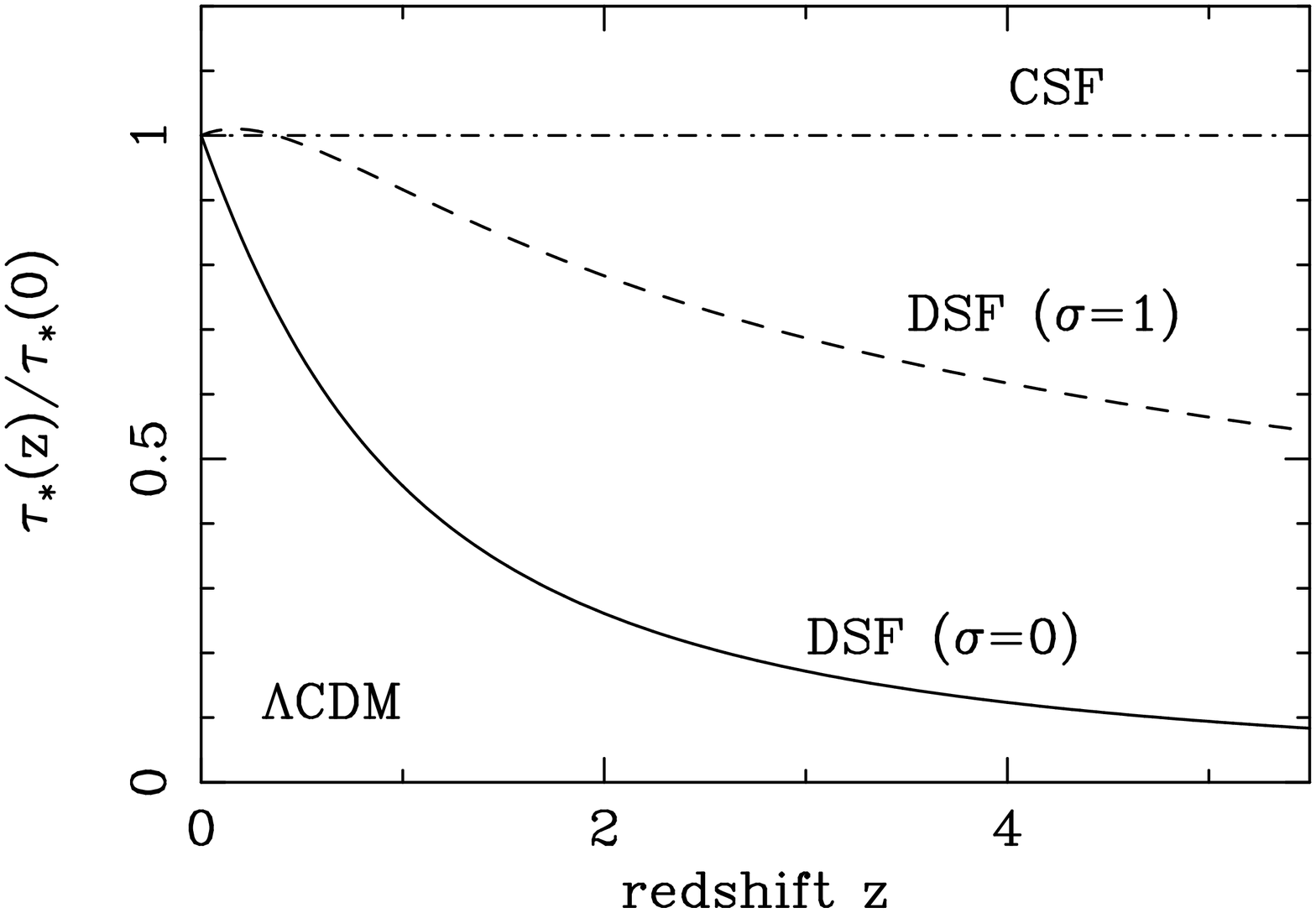}

\caption{ Star formation timescales $\tau_{*}$ as a function of redshift $z$
in a $\Lambda$-dominated flat CDM model.  The timescale is normalized 
by that evaluated at present.  The dot-dashed line denotes the constant SF 
model (CSF).  The solid line denotes the dynamical star formation model 
(DSF) with $\sigma=0$ for which the timescale is proportional to the 
dynamical timescale of disk.  The dashed line denotes the DSF with 
$\sigma=1$ as an intermediate case between CSF and DSF with $\sigma=0$.  
}
\label{fig:sfr}
\end{inlinefigure}

Massive stars explode as Type II supernovae (SNe) and heat up the
surrounding cold gas.  This SN feedback reheats the cold gas at a rate
of $\dot{M}_{\rm reheat}={M_{\rm cold}}/{\tau_{\rm reheat}}$, where the
timescale of reheating is given by

\begin{equation}
\tau_{\rm reheat}=\left(\frac{V_{\rm circ}}{V_{\rm hot}}
\right)^{\alpha_{\rm hot}} \tau_{*}.
\end{equation}

The free parameters $V_{\rm hot}$ and $\alpha_{\rm hot}$ are determined
by matching the local luminosity function of galaxies with observations.

With the parameters $\dot{M}_{*}$ and $\dot{M}_{\rm reheat}$ thus
determined, we obtain the masses of hot gas, cold gas, and disk stars as
a function of time during the evolution of galaxies.  Chemical
enrichment is also taken into account, adopting {\it heavy-element
yield} of $y=0.038=2Z_{\odot}$, but changing this value of $y$ has a
minimal effect on the results described below.

When two or more progenitor halos have merged, the newly formed larger
halo should contain at least two or more galaxies which had originally
resided in the individual progenitor halos.  By definition, we identify
the central galaxy in the new common halo with the central galaxy
contained in the most massive one of the progenitor halos.  Other
galaxies are regarded as satellite galaxies.  These satellites merge by
either dynamical friction or random collision.  The timescale of merging
by dynamical friction $\tau_{\rm fric}$ is given by \citet{bt87}, which
is estimated from the baryonic mass of satellite.  When the time elapsed
after merging of progenitor halos exceeds $\tau_{\rm fric}$, a satellite
galaxy in the common halo is accreted to the central galaxy.  On the
other hand, the mean free timescale of random collision $\tau_{\rm
coll}$ is given by \citet{mh97}.  With a probability $\Delta t/\tau_{\rm
coll}$, where $\Delta t$ is the time step corresponding to the redshift
interval $\Delta z$ of merger tree of dark halos, a satellite galaxy
merges with another satellite picked out randomly.

\begin{figure}
\plotone{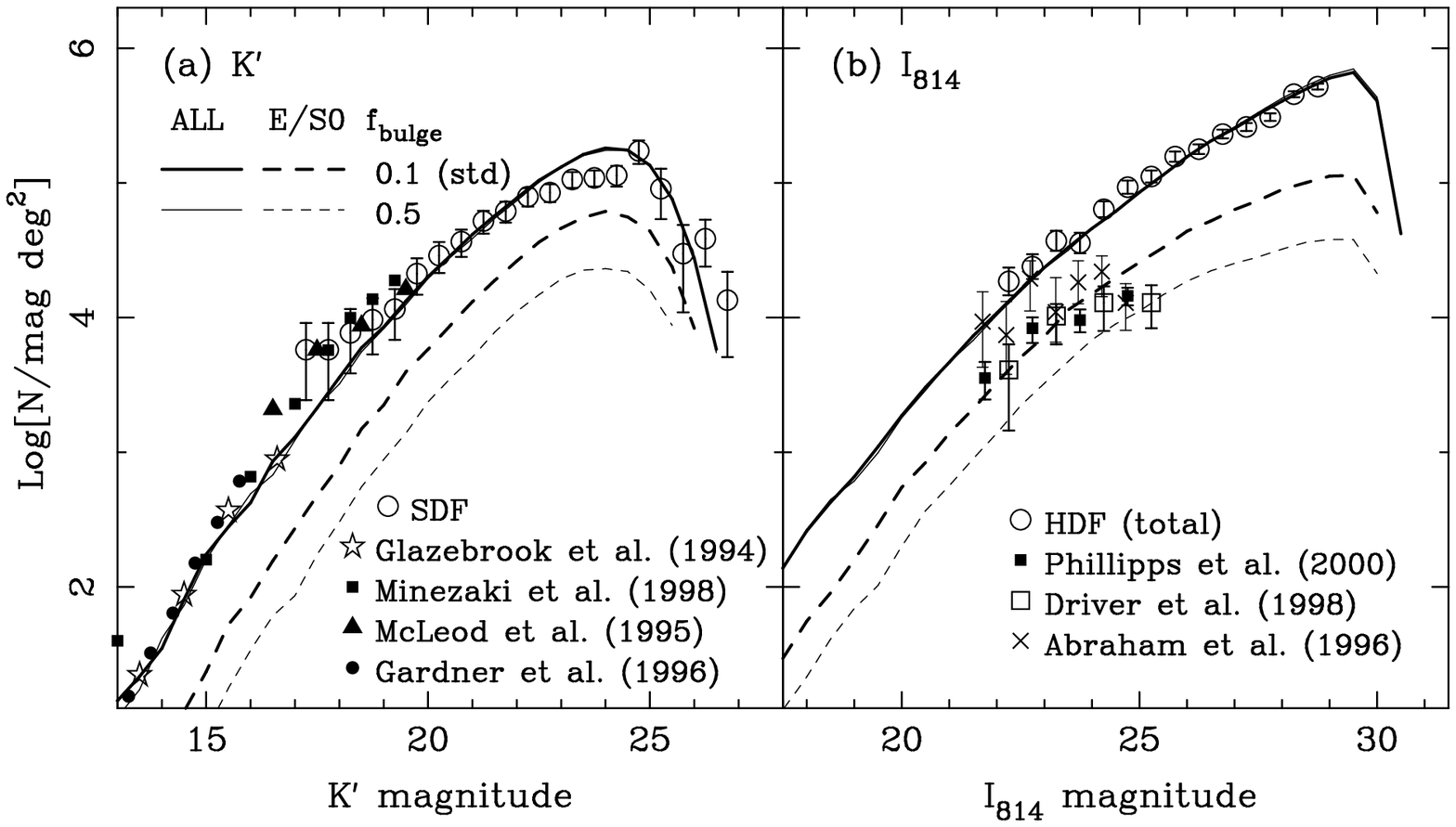}
\caption{Morphology-dependent number-magnitude relations in the LC
model.  The solid and dashed lines indicate the total and early-type
(E/S0) galaxy counts, respectively, in the LC model.  The thick lines of
these two are for $f_{\rm bulge}=0.1$, and the thin lines for $f_{\rm
bulge}=0.5$ model.  }
\label{fig:sdf_morph}
\end{figure}

Consider the case when two galaxies of masses $m_1$ and $m_2 (>m_1)$
merge together.  If the mass ratio $f=m_1/m_2$ is larger than a certain
critical value of $f_{\rm bulge}$, we assume that a starburst occurs and
that all of the cold gas turns into stars and hot gas, which fills the
resulting halo, and all of the stars populate the bulge of a new galaxy.
On the other hand, if $f<f_{\rm bulge}$, no starburst occurs, and a
smaller galaxy is simply absorbed into the disk of a larger galaxy.
This division into major and minor modes is only for simplicity and we
adopt $f_{\rm bulge}=0.1$ as a standard value.  As shown in Figure
\ref{fig:sdf_morph}, the dependence of total counts on $f_{\rm bulge}$
is negligible (see \S\ref{sec:morph}).  Although there is a model to
treat continuous burst activity, say, as a function of the mass ratio
$f$ \citep{spf01}, we adopt the above simple division because of the
negligible denpendence on $f_{\rm bulge}$.

\subsection{Photometric Properties of Galaxies}

The above processes are repeated until the output redshift and then the
SF history of each galaxy is obtained.  For the purpose of comparison
with observation, we use a stellar population synthesis approach, from
which the luminosities and colors of model galaxies are calculated.
Given the SFR as a function of time or redshift, the absolute luminosity
and colors of individual galaxies are calculated using a population
synthesis code by \citet{ka97}. The stellar metallicity grids in the
code cover a range from $Z_{*}=$0.0001 to 0.05. Note that we now define
the metallicity as mass fraction of metals, for example, the solar
metallicity is 0.019.  The initial stellar mass function (IMF) that we
adopt is the power-law IMF of Salpeter form, with lower and upper mass
limits of $0.1M_{\odot}$ and $60M_{\odot}$, respectively.  Then,
following Cole et al. (1994), we introduce a parameter defined as
$\Upsilon=(M_{\rm lum}+M_{\rm BD})/M_{\rm lum}$, where $M_{\rm lum}$ is
the total mass of luminous stars with $m\geq 0.1M_\odot$ and $M_{\rm
BD}$ is that of invisible brown dwarfs.

Our model of estimating the optical depth of internal dust has been
improved over that adopted by NTGY.  We take the usual assumption that
the abundance of dust is proportional to the metallicity of cold gas,
and that the optical depth is proportional to the column density of
metallicity.  Then the optical depth $\tau$ is given by
\begin{equation}
 \tau\propto\frac{M_{\rm cold}Z_{\rm cold}}{r_{e}^{2}}(1+z)^{-\gamma},
\label{eqn:dust}
\end{equation}
where $r_{e}$ is the effective radius of the galactic disk, described in
the next subsection.  The proportionality constant is determined by
matching the extinction, $A_{V}\approx 0.2$, for galaxies typical of the
Milky Way, assuming the slab dust model according to our previous paper
(NTGY) and \citet{sp99}.  We adjust the high redshift properties of
dust-to-metal ratio and dust clumpiness by changing $\gamma$.  As a
standard model, we simply adopt $\gamma=0$.  It will be also
investigated in \S\ref{sec:udust} how the galaxy counts are affected by
$\gamma$.

For some classes of galaxies, the screen dust model has been suggested
to be a good approximation, rather than the slab dust model.  For
example, some galaxies such as nearby starburst galaxies and hyper
extremely red objects (HEROs) are very red and therefore their dust
distribution should be approximated by the screen model \citep{tyimm}.
Besides recent analysis of source counts in far-infrared and
submilimeter wavelengths also favors the screen dust model \citep{tt}.
In this paper, however, we adopt the slab model as a standard one
because main contribution in UV/optical and near-infrared passbands is
considered to be galaxies which are not heavily extincted under modest
star formation activity.  Of course, even for these normal galaxies the
screen dust model might be a good approximation.  Therefore we will
check the screen dust model in \S\ref{sec:udust}.  While we check only
these two models, the uncertainty cased by the dust model would be
clarified by checking these models and by varying $\gamma$.

Emitted light from distant galaxies is absorbed by Lyman lines and Lyman
continuum in intervening intergalactic \ion{H}{1} clouds.  We used an
optical depth calculated by \citet{yp94} to account for this.

We classify galaxies into different morphological types according to the
$B$-band bulge-to-disk luminosity ratio $B/D$.  In this paper, following
\citet{sdv86}, galaxies with $B/D\geq 1.52$, $0.68\leq B/D<1.52$, and
$B/D<0.68$ are classified as ellipticals, S0s, and spirals,
respectively.  \citet{kwg93} and \citet{bcf96} showed that this method
of type classification reproduces the observed type mix well.

We then assess whether the surface brightnesses of model galaxies are
above the detection thresholds of the SDF and HDF observations.  The
selection effects in our SAM analysis are evaluated as follows.  Using
the intrinsic size of model galaxies, described in the next subsection,
their surface brightness profiles in the observers frame are given by a
convolution of the luminosity profile with a Gaussian point-spread
function.  Here we assume an exponential profile for spiral galaxies,
and a de Vaucouleurs profile for elliptical and lenticular galaxies.  We
note that model galaxies with surface brightness higher than the
threshold $S_{\rm th}$, and isophotal areas larger than the minimum area
$A_{\rm th}$ are actually detected as galaxies \citep{y93}.  Following
\citet{ty00} and \citet{t01}, we adopt $S_{\rm th}=25.59$
mag~arcsec$^{-2}$ in $K'$, $S_{\rm th}=24.10$ mag~arcsec$^{-2}$ in $J$,
and $A_{\rm th}=$0.24 arcsec$^{2}$ for the SDF galaxies, and $S_{\rm
th}=27.5$ mag~arcsec$^{-2}$ in $V_{606}$, $S_{\rm th}=27.0$
mag~arcsec$^{-2}$ in $I_{814}, U_{300}$ and $B_{450}$, and $A_{\rm
th}=$0.04 arcsec$^{2}$ for the HDF galaxies.

For the SDF galaxies, we also take into account the detection
probability, that is, completeness.  Noise and statistical fluctuations
in the data prevent complete source detections at any specified limits
of surface brightness and size.  \citet{t01} found that the dispersion
of observed isophotal area around its true value can be fitted by

\begin{equation}
 \sigma_{A}(m, d_{\rm ob})=c(A_{1}-A_{2})^{1/2}d_{\rm ob},
\end{equation}
where $m$ is the total magnitude of an object, $d_{\rm ob}$ is the FWHM
size, and $A_{1}$ and $A_{2}$ are the isophotal areas corresponding to
the isophotal level 0.8 and 1.2 times brighter than $S_{\rm th}$,
respectively.  Assuming a Gaussian distribution of observed isophotal
area with this dispersion, we obtain the probability that the observed
isophotal area is larger than the threshold value $A_{\rm th}$.
Additional details of these selection effects are described in
\citet{ty00} and \citet{t01}.

\subsection{Galaxy Size}\label{sec:ml}

We assume that the size of spiral galaxies is determined by a radius at
which the gas is supported by rotation, under the conservation of
specific angular momentum of hot gas that cools and contracts.  We also
assume that the initial specific angular momentum of the gas is the same
as that of the host dark halo.  Acquisition of the angular momentum of
dark halos is determined by the tidal torques in the initial density
fluctuation field \citep{w84, ct96a, ct96b, ng97}.  The distribution of
the dimensionless spin parameter $\lambda_{\rm H}$ is well approximated
by a log-normal distribution \citep{mmw98},
\begin{equation}
 p(\lambda_{\rm H})d\lambda_{\rm H}=
\frac{1}{\sqrt{2\pi}\sigma_{\lambda}}
\exp\left[-\frac{(\ln\lambda_{\rm H}-\ln\bar{\lambda})^2}
{2\sigma_{\lambda}^{2}}\right] d\ln\lambda_{\rm H},
\end{equation}
where $\bar{\lambda}$ is the mean value of spin parameter and
$\sigma_{\lambda}$ is its logarithmic variance. We adopt
$\bar{\lambda}=0.05$ and $\sigma_{\lambda}=0.5$.  (Note that in our
previous paper we adopted $\lambda_{\rm H}=0.05$ for all spirals.)  If
the specific angular momentum is conserved, the effective radius $r_{e}$
of a presently observed galaxy at $z=0$ is related to the initial radius
$r_{i}$ of the progenitor gas sphere via
$r_{e}=(1.68/\sqrt{2})\lambda_{H}r_{i}$ \citep{f79, fe80, f83}.  The
initial radius $r_i$ is set to be the smaller one between the virial
radius of the host halo and the cooling radius.  A disk of a galaxy
grows due to cooling and accretion of hot gas from more distant envelope
of its host halo.  In our model, when the estimated radius by the above
equation becomes larger than that in the previous step, the radius grows
to the new larger value in the next step.

Size estimation of high-redshift spiral galaxies, however, carries
uncertainties because of the large dispersion in their observed size
distribution.  Allowing for the possibility that the conservation of
angular momentum is not complete, we generalize this size estimation by
introducing a simple redshift dependence,
\begin{equation}
 r_{e}=\frac{1.68}{\sqrt{2}}\lambda_{H}r_{i}(1+z)^{\rho},
\end{equation}
where $\rho$ is a free parameter; we simply use $\rho=0$ as a reference
value in this paper.  Results for other values of $\rho$ will be given
in \S\ref{sec:uncertainties}.  The effect of changing $\rho$ emerges in
the selection effects due to the cosmological dimming of surface
brightness and in the dust extinction because the dust column density
also changes with galaxy size.

Figure \ref{fig:rad} shows the effective disk radii of spiral galaxies
at $z=0$ as a function of their luminosity for the models of SC (thick
solid line), OC (dashed line), and LC (dot-dashed line).  The adopted
parameters in these models are tabulated in Table \ref{tab:astro}.  All
the models well reproduce the observed disk size-magnitude relation
(thin solid line) compiled by \citet{ty00} based on the data taken from
\citet{isib96}. The redshift dependence of such relation through
changing $\rho$ in equation 6 will be discussed in
\S\ref{sec:uncertainties}.

The sizes of early-type galaxies, which likely form from galaxy mergers,
are determined by the virial radius of the baryonic component. When a
major merger of galaxies occurs, the merged system consisting of stars
and cold gas initially virializes to satisfy
\begin{equation}
 V_{\rm circ}^{2}=\frac{G(M_{*}+M_{\rm cold})}{r_i}.
\end{equation}
In this paper we take into account the dynamical response to the
``shallowing'' of the gravitational potential due to the mass loss
caused by the SN feedback.  Then, after the initial virialization, the
system expands while losing the cold gas, with the adiabatic invariant
$(M_*+M_{\rm cold})r$ kept constant (e.g., Yoshii \& Arimoto 1987).
Then the final radius of the system, consisting only of stars after the
loss of the cold gas, is given by
\begin{equation}
 r_{f}=\frac{M_{*,i}+M_{\rm cold}}{M_{*,f}}r_{i},
\end{equation}
where $M_{*,i}$ and $M_{*,f}$ are the total masses of stars in the
system before and after the mass loss, respectively.  The effect of the
dynamical response is most prominent for dwarf galaxies of lower
circular velocity, and can explain the properties of local dwarf
ellipticals, which will be presented in a separate paper.  However, this
effect has only a minor impact on our predicted galaxy number counts,
because luminous galaxies of larger circular velocity are the main
contributors to galaxy number counts at the faint limits under
consideration. In order to match with the observed effective radius, we
here introduce a scaling parameter, $f_{\rm b}$, such that $r_e=f_{\rm
b}r_f$. The adopted values of $f_{\rm b}$ for SC, OC, and LC are
tabulated in Table \ref{tab:astro}.

\begin{inlinefigure}
\includegraphics[width=\hsize]{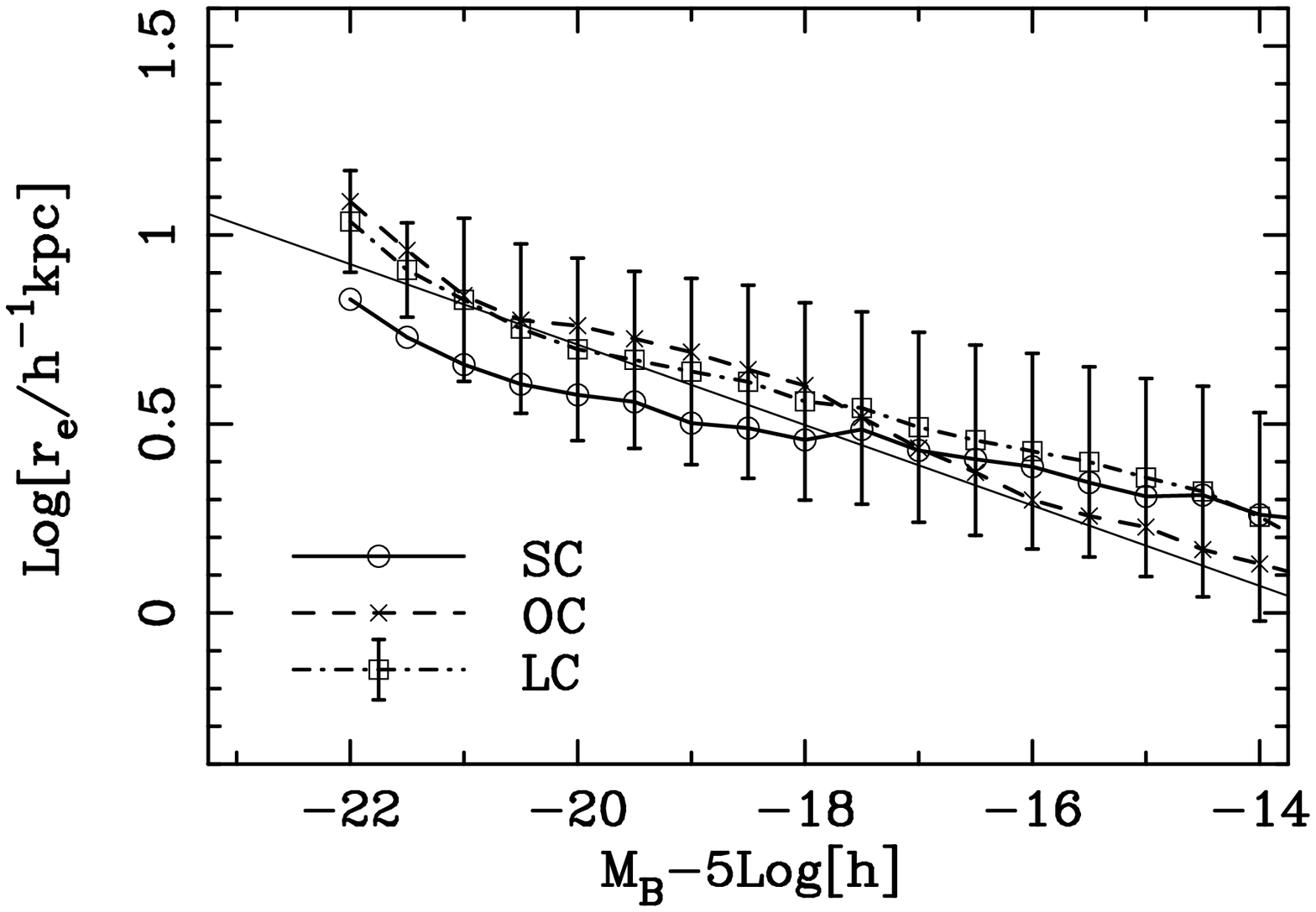}

\caption{Galaxy sizes.  The thick solid line connecting open circles, 
the dashed line connecting crosses, and the dot-dashed line connecting open 
squares show theoretical results for effective radii of late-type 
galaxies in the SC, OC, and LC models, respectively.  We show 1$\sigma$ 
scatter in the LC model by error bars.  Similar scatter in other models 
are not shown.  The thin line shows the observed best-fit relation given 
by \citet{ty00} for spiral galaxies, based on the data taken from Impey et 
al. (1996).  
}
\label{fig:rad}
\end{inlinefigure}

\section{PARAMETER SETTINGS}\label{sec:set}

We normalize the model parameters so as to agree with various local
observations.  As shown in NTGY, this normalization procedure is an
essential ingredient in the analysis of galaxy counts. 

The cosmological parameters ($\Omega_0$, $\Omega_\Lambda$, $h$,
$\sigma_8$) adopted in this paper are tabulated in Table
\ref{tab:astro}.  For all the models (SC, OC, and LC), the baryon
density parameter $\Omega_{\rm b}=0.015h^{-2}$ is used in common.  For
OC and LC the value of $\sigma_8=1$ is determined from observed cluster
abundances \citep{ecf96}.

The astrophysical parameters ($V_{\rm hot}$, $\alpha_{\rm hot}$,
$\tau_*^0$, $\alpha_*$, $f_{\rm b}$, $f_{\rm bulge}$, $\Upsilon$) are
constrained from local observations as discussed below.  The adopted
values are almost the same as in our previous paper.  Slight differences
have resulted from modifications to the model, such as the improved
scheme of dust extinction.  In this paper we only show the result of CSF
(equation 1) as a standard, and other parameters used in common,
including $\gamma=0$ (equation 3), $\rho=0$ (equation 6), and
($\bar{\lambda}, \sigma_{\lambda}$)=(0.05, 0.5) (equation 5).

\begin{inlinefigure}
\includegraphics[width=\hsize]{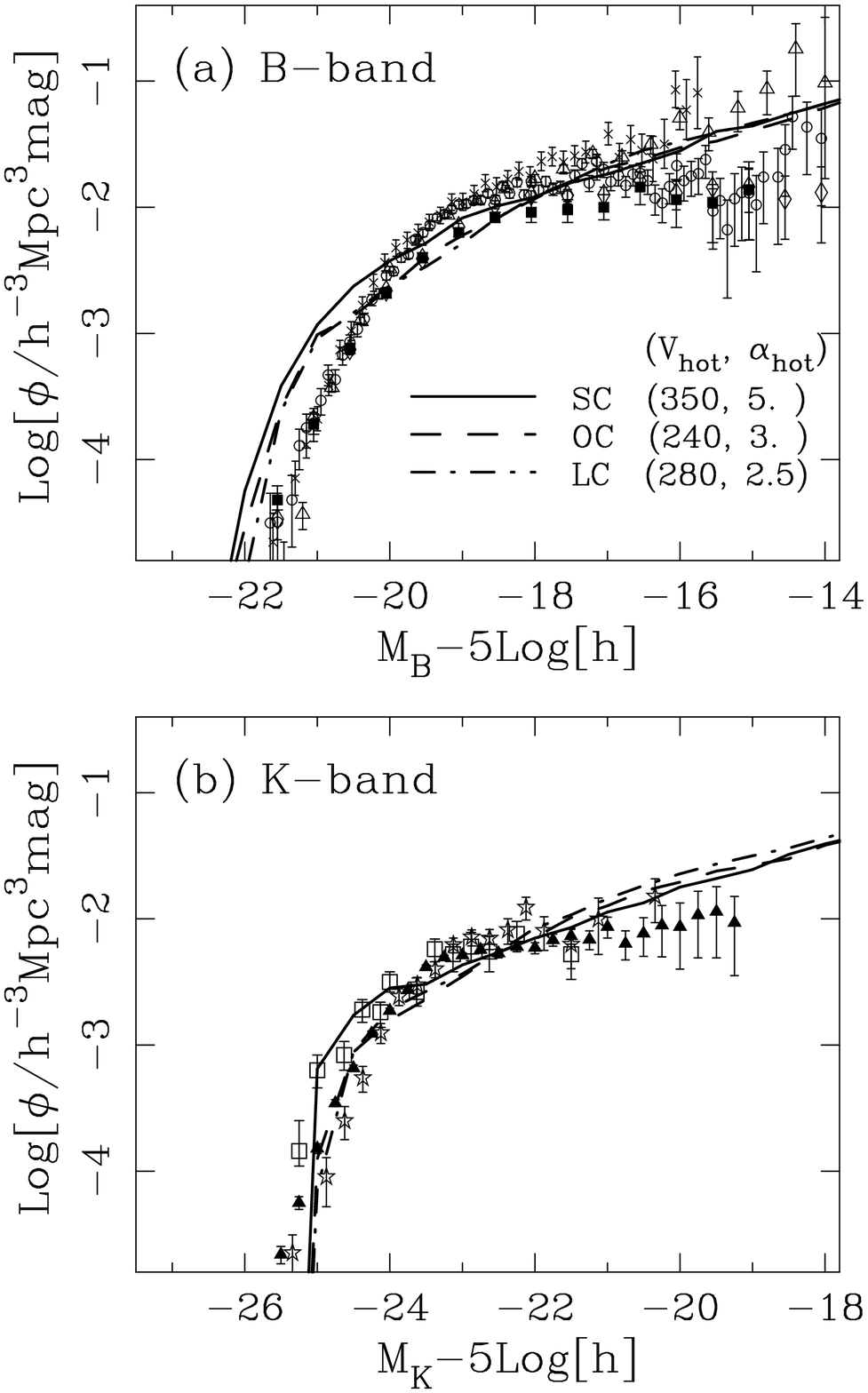}

\caption{Local luminosity functions in the (a) $B$ band and (b) $K$
band.  The solid, dashed, and  dot-dashed lines indicate the SC,
OC, and LC models, respectively.  Symbols with error bars in (a)
indicate the observational data from APM (Loveday et al. 1992, {\it
filled squares}), ESP (Zucca et al. 1997, {\it open triangles}),
Durham/UKST (Ratcliffe et al. 1998, {\it open diamonds}), 2dF (Folkes
et al. 1999, {\it open circles}), and SDSS (Blanton et al. 2000, {\it
crosses}).  Note that the SDSS luminosity function shown here is that
with the same detection limit as employed in the 2dF survey.  Symbols
in (b) indicate the data from Mobasher et al.  (1993, {\it open
squares}), Gardner et al. (1997, {\it open stars}), and 2MASS (Cole et
al. 2000, {\it filled triangles}).  
}
\label{fig:lf}
\end{inlinefigure}

First, the SN feedback-related parameters ($V_{\rm hot}, \alpha_{\rm
hot}$) and the mass fraction $\Upsilon$ of invisible stars are almost
uniquely determined if their values are chosen so as to reproduce the
local luminosity function.  Figure \ref{fig:lf} shows theoretical
results for the models of SC (thick solid line), OC (dashed line), and
LC (dot-dashed line), which are tabulated in Table \ref{tab:astro}.
Symbols with error bars indicate observational results from the $B$-band
redshift surveys, such as APM \citep{l92}, ESP \citep{z97}, Durham/UKST
\citep{r98}, 2dF \citep{f99} and SDSS \citep{b01}, and from the $K$-band
redshift surveys (Mobasher et al. 1993; Gardner et al. 1997; 2MASS, Cole
et al. 2001).  As done in NTGY, we regard the APM result as a standard
one.  This gives somewhat less normalization of luminosity function
compared with other surveys', probably because of its worse completeness
compared to the other recent surveys.  Nevertheless, in this paper, we
determine the SN feedback-related parameters in the above way because
the dependence of these parameters on galaxy counts has been already
investigated by NTGY and then because it has been found that the shape
of counts is not significantly affected.  We checked that the shape of
the counts in the $K'$-band are also not affected by changing the
luminosity function significantly.

Next, the SFR-related parameters ($\tau_*^0$, $\alpha_*$) are determined by
using the cold gas mass fraction in late-type galaxies.  The gas fraction
depends on both the SN feedback-related parameters and on the SFR-related ones.
The former parameters determine the gas fraction expelled from galaxies and
the latter ones the gas fraction that is converted into stars.  Therefore, in
advance of determining the SFR-related parameters, the SN feedback-related
parameters must be determined by matching the local luminosity function.

Figure \ref{fig:gas} shows the ratio of cold gas mass relative to
$B$-band luminosity of spiral galaxies as a function of their luminosity.  
Theoretical results are shown for the models of SC (thick solid line), 
OC (dashed line), and LC (dot-dashed line).  We here assume that 75\% of
the cold gas in these models is comprised of hydrogen, i.e., $M_{\rm
HI}=0.75M_{\rm cold}$.  \ion{H}{1} data, taken from \citet{hr88}, are
shown by open squares with error bars.  Since their data do not include the 
fraction of H$_{2}$ molecules, the observational result should be regarded
as providing a lower limit of the cold gas mass fraction.

\begin{inlinefigure}
\includegraphics[width=\hsize]{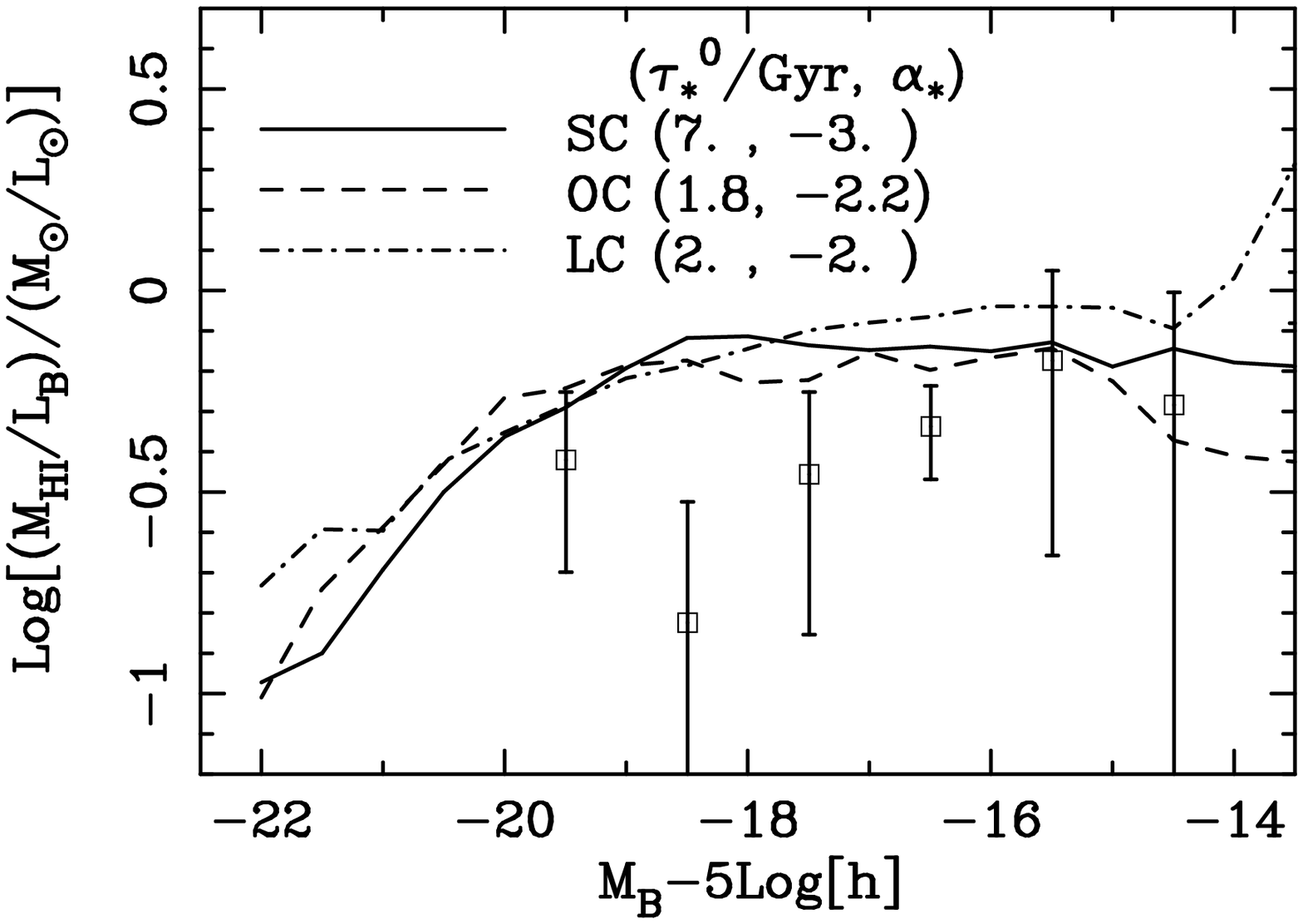}

\caption{Cold gas mass relative to $B$-band luminosity of spiral
galaxies.  The solid, dashed, and dot-dashed curves show the SC, OC,
and LC models, respectively.  The open squares indicate the
observational data for atomic neutral hydrogen taken from Huchtmeier \&
Richter (1988).  In the models the cold gas consists of all species of
elements, therefore its mass is multiplied by 0.75, i.e., 
$M_{\rm HI}=0.75M_{\rm cold}$, which corresponds to the fraction of 
hydrogen.  Because the observational data denote only atomic hydrogen, 
they should be regarded as lower limits of the ratio.  
}
 \label{fig:gas}
\end{inlinefigure}

\section{RESULTS}

\subsection{Galaxy Number Counts}\label{sec:counts}

\begin{figure}
\plotone{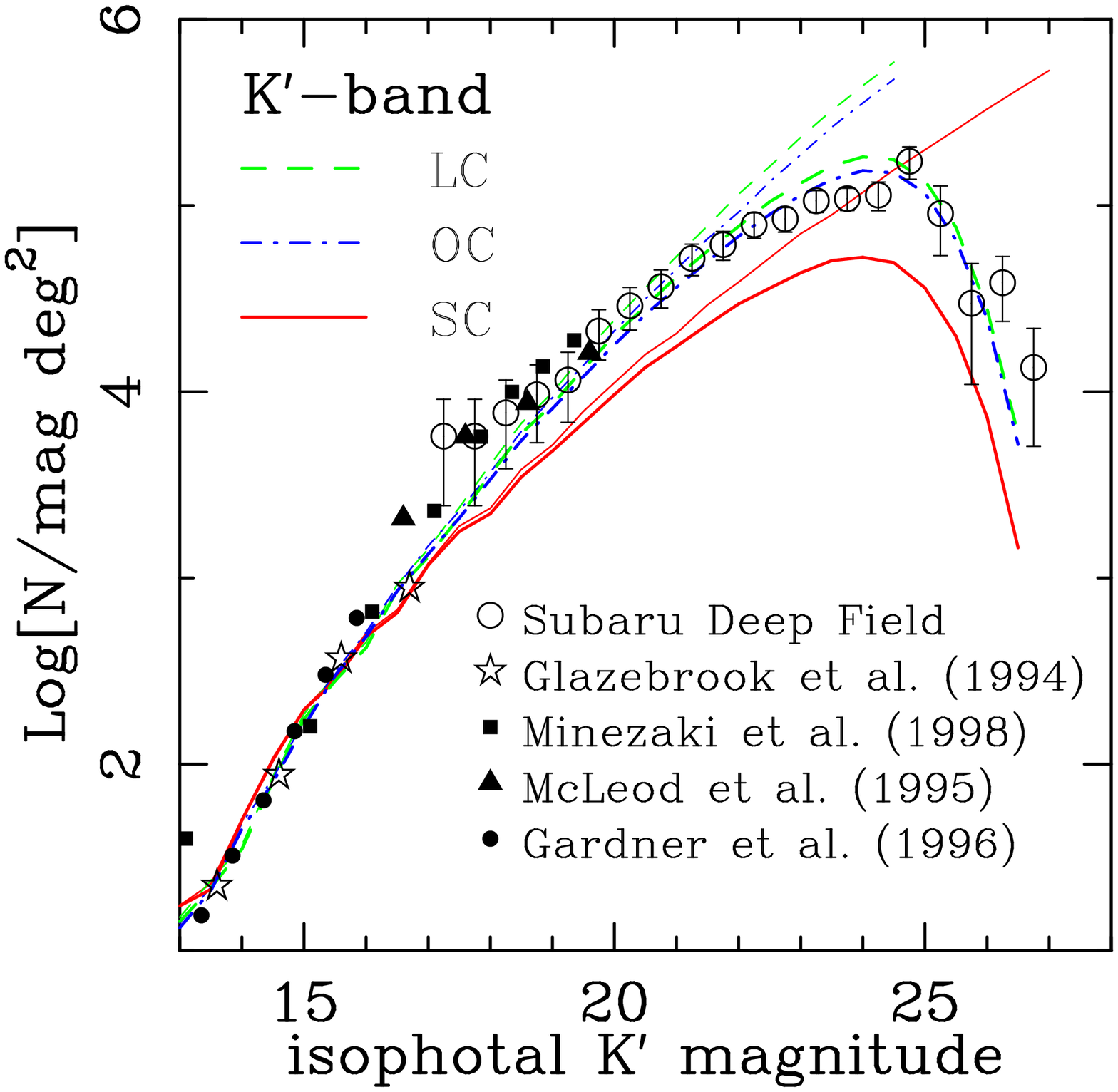}
\caption{Number-magnitude relations for various cosmological models in
the Subaru $K'$-band.  The solid, dot-dashed, and dashed lines indicate
SC, OC, and LC, respectively.  The thick lines denote the models including
the selection effects, while the thin lines denote those without the
selection effects.  The open circles with errorbars indicate the raw counts 
in the SDF.  The other symbols indicate other observational data referred
in the figure.  
}
\label{fig:counts_sdf}
\end{figure}

Figure \ref{fig:counts_sdf} shows the galaxy number counts in the Subaru
$K'$-band as a function of apparent isophotal $K'$ magnitude for the
models of SC, OC, and LC.  Thick lines show theoretical predictions
based on the observational conditions, including the selection effects
from the cosmological dimming of surface brightness and completeness for
galaxies in the SDF.  Open circles indicate the observed SDF raw counts
in isophotal magnitude.  The number of galaxies in the SC model is too
small to explain the observed raw counts (open circles with error bars).
On the other hand, the OC and LC models explain the SDF counts equally
well.  For reference, the data from other surveys are also shown by the
specified symbols, after applying a transformation of $K'=K+0.1$.

Thin lines in this figure are theoretical predictions without the selection
effects. It is evident that the effects become important at $K'> 20$, and make
the count slope turn over at $K'\sim 24$, where the observed raw counts reach
a maximum.  The predicted counts with and without the selection effects
differ by a factor of three at $K'\sim 24$, which is much greater than the
observational error.  Thus, incorporation of the selection effects is essential
in the SAM analysis of the galaxy number counts.

\begin{figure}
\plotone{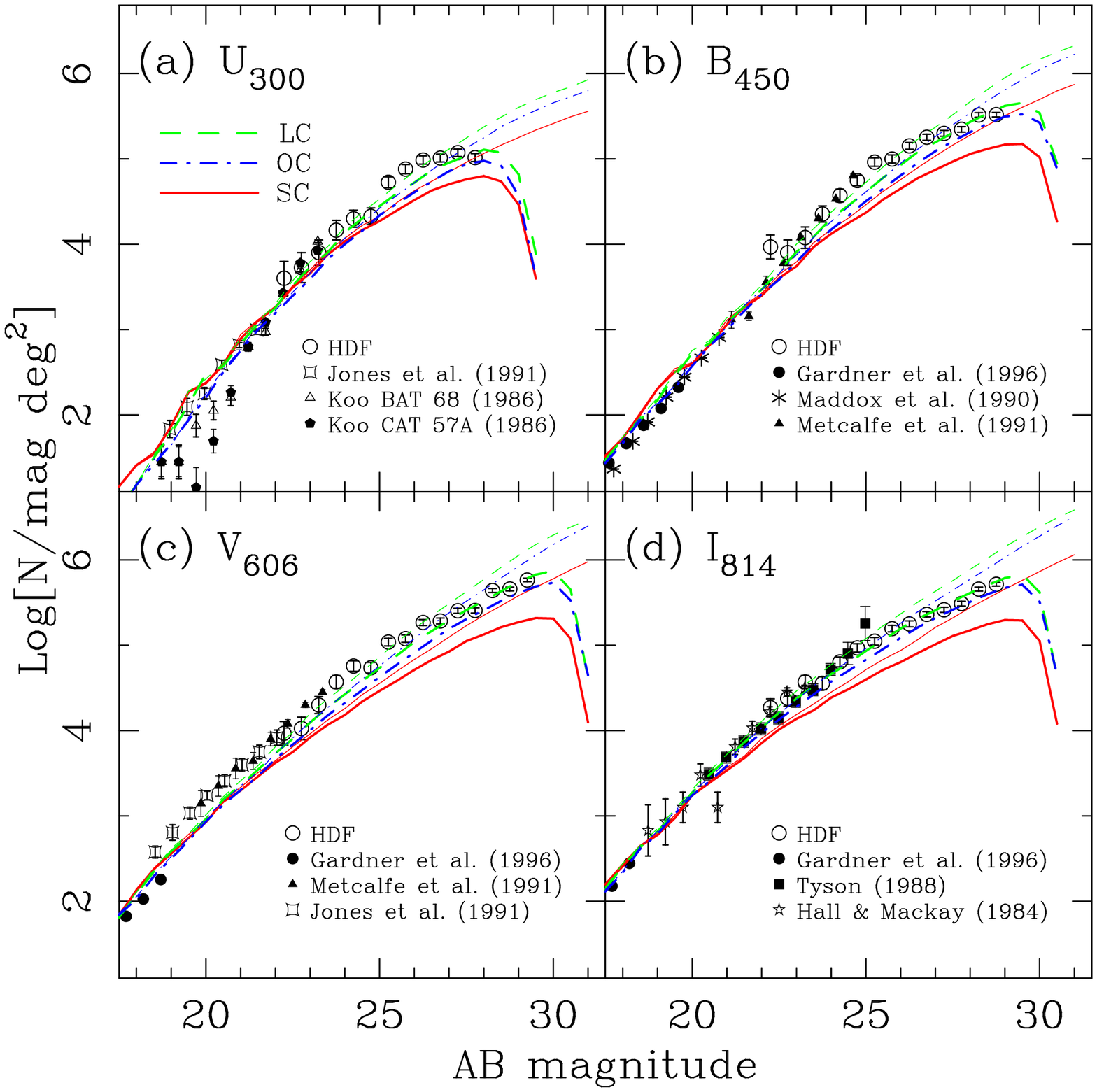}
\caption{Number-magnitude relations for various cosmological models in
the {\it HST} $UBVI$-bands.  Types of lines are the same as Figure
\ref{fig:counts_sdf}.  The symbols indicate observational data referred to
in the figure.  
}
\label{fig:counts_hdf}
\end{figure}

We check for consistency with the HDF counts in the $UBVI$ bands in
Figure \ref{fig:counts_hdf}.  The types of lines are the same as in
Figure \ref{fig:counts_sdf}.  We see that the OC and LC models, which
reproduce the near-infrared SDF counts, give a better agreement with the
UV/optical HDF counts when compared with our previous models in NTGY.
On the other hand, the SC model significantly falls below those observed
in all the $UBVIK'$ bands.  Thus we conclude that our SAMs well
reproduce the multiband galaxy number counts in a low-density universes
with or without a cosmological constant.

\subsection{Isophotal Area-Magnitude Relation}\label{sec:angular}

\begin{figure}
\plotone{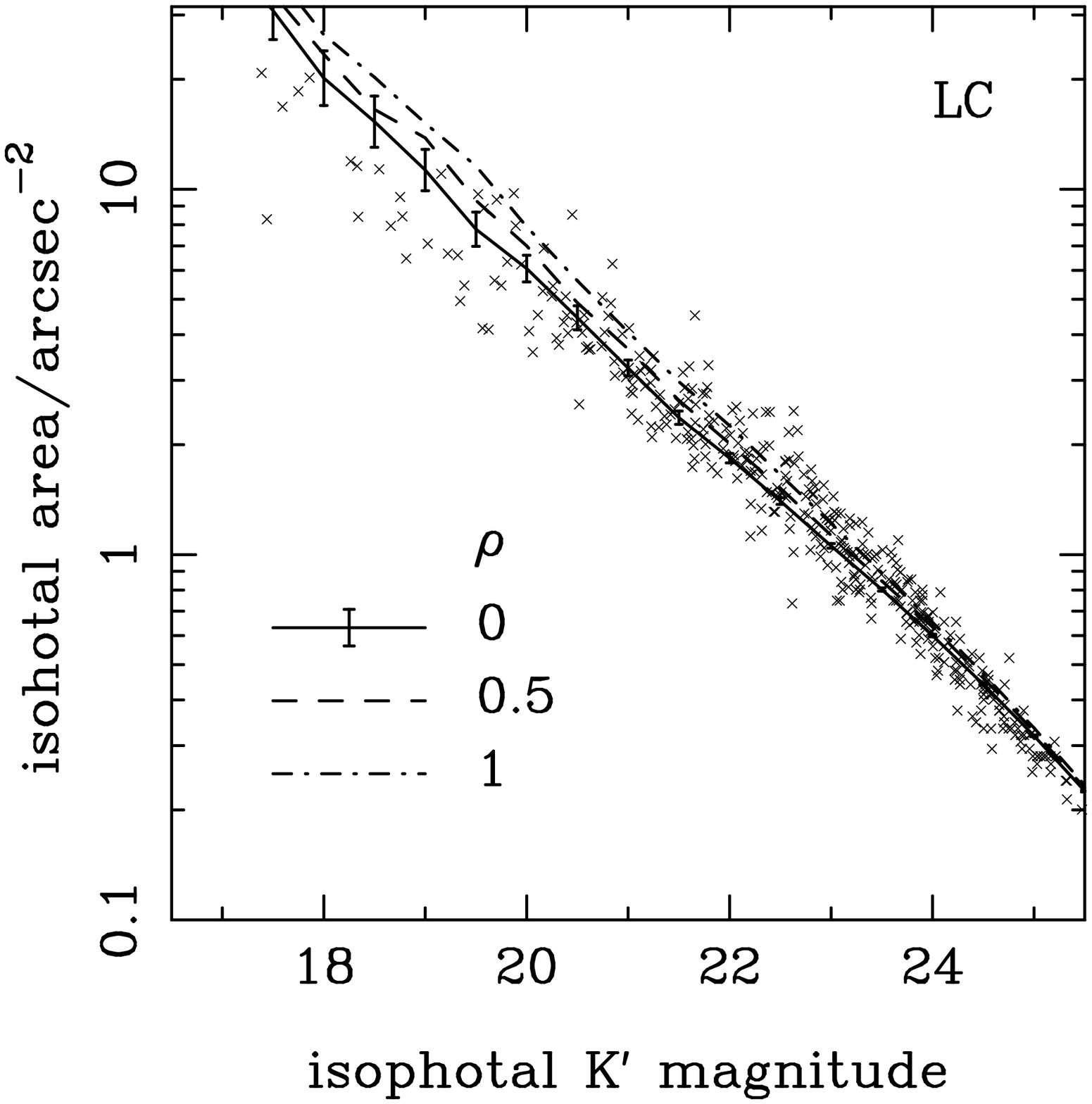}
\caption{Isophotal area of galaxies detected in both the $K'$ and $J$
bands in the LC model.  The parameter $\rho$ is varied from 0 (solid
line), via 0.5 (dashed line), to 1 (dot-dashed line).  The 1$\sigma$
scatters in the model with $\rho=0$ are shown by error bars. The crosses
indicate the data in the SDF.  }
 \label{fig:angsize}
\end{figure}

As stressed in the previous section and in NTGY, the selection effects
from the cosmological dimming of surface brightness of galaxies cannot
be ignored in the SAM analysis of galaxy number counts.  This indicates
that the size of high-redshift galaxies must be modeled properly.
Figure \ref{fig:angsize} plots the SDF galaxies in the isophotal
area-magnitude diagram.  The data plotted are those for the SDF galaxies
that are detected in both the $K'$ and $J$ bands.  Only the LC model is
shown, with various values of $\rho=0$ (solid line), 0.5 (dashed line),
and 1 (dot-dashed line), because other parameters involving the adopted
cosmology and dust extinction hardly affect the result.  The isophotal
area becomes smaller for fainter apparent magnitude; at faint limits it
reaches the minimum for detection.  All predictions with $\rho=0-1$ give
a convergent result at faint limit of $K'\sim 24$. This convergence
occurs because galaxies with larger areas at $K'\sim 24$ have surface
brightnesses below the threshold and remain undetected.  We find from
this figure that our SAM galaxies well reproduce the observed
area-magnitude relation, and are consistent with the SDF galaxies, only
when the selection bias against faint galaxies with high redshift and/or
low surface brightness is taken into account in the analysis.

\subsection{Redshift Distributions}

\begin{figure}
\plotone{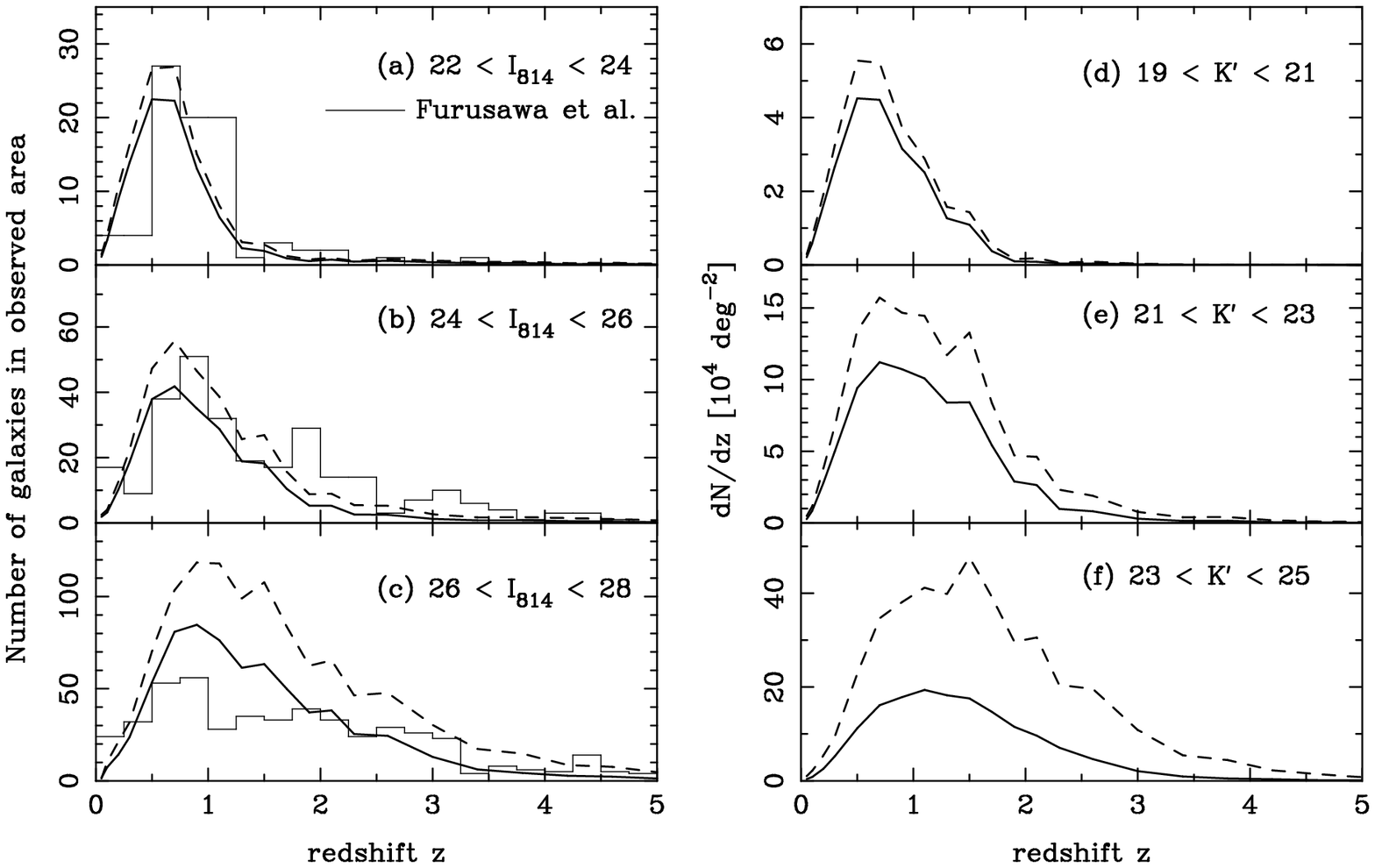}
\caption{Redshift distributions of galaxies for a $\Lambda$-dominated flat 
CDM model.  Left panels: the $I_{814}$-band.  Right panels: the $K'$-band.  
The solid and dashed lines denote the LC model with and without the selection 
effects, respectively.  The histograms in the left panels indicate the 
redshift distributions of the HDF galaxies derived by \citet{f00}, using an 
improved technique of photometric redshift estimation.
}
\label{fig:zdist}
\end{figure}

The left panels in Figure \ref{fig:zdist} show the redshift distributions 
for the HDF galaxies in the $I_{814}$ band.  The solid and dashed lines are 
theoretical predictions for the LC model with and without the selection 
effects, respectively.  The histograms are the observed photometric redshift 
distributions given by \citet{f00}.  As sown in NTGY, the LC model well 
reproduces the observed redshift distribution as a function of apparent 
magnitude.  

The right panels of Figure \ref{fig:zdist} show the $K'$-band
predictions for the LC model.  Compared with \citet{t01}, in which the
traditional models of galaxy evolution are used, the peak location for
our redshift distribution moves to lower redshift.  Therefore, redshifts
of the SDF galaxies, if measured by either a spectroscopic or
photometric method, give an important insight into physical processes of
galaxy formation and evolution.

\section{MODEL UNCERTAINTIES AT HIGH REDSHIFT}\label{sec:uncertainties}

In this section we discuss uncertainties in predicting the number counts
of galaxies.  Such uncertainties arise mainly from the SF timescale
(\S\ref{sec:usf}), the galaxy size (\S\ref{sec:usize}), and the dust
extinction (\S\ref{sec:udust}).  As for the SN feedback-related
parameters, we find that the dependence of our result on their values
has a similar tendency to NTGY, so that our conclusion of preferring a
low-density universe is not altered by uncertainty associated with these
parameters.  Finally we will discuss the morphological counts
(\S\ref{sec:morph}).

\subsection{Star Formation Timescale}\label{sec:usf}

We evaluate the effects of changing the SF timescale from CSF to DSF
($\sigma=0$ or 1) in the $\Lambda$CDM model (see equation 1).  Note that CSF in
the EdS universe is equivalent to DSF with $\sigma=1.5$, thus the case of DSF
with $\sigma=1$ is in between CSF and DSF with $\sigma=0$, as shown in Figure
\ref{fig:sfr}. Hereafter, the two cases of DSF with $\sigma=0$ and 1 in the
$\Lambda$CDM model are referred to as LD0 and LD1, respectively.

\begin{figure}
\plotone{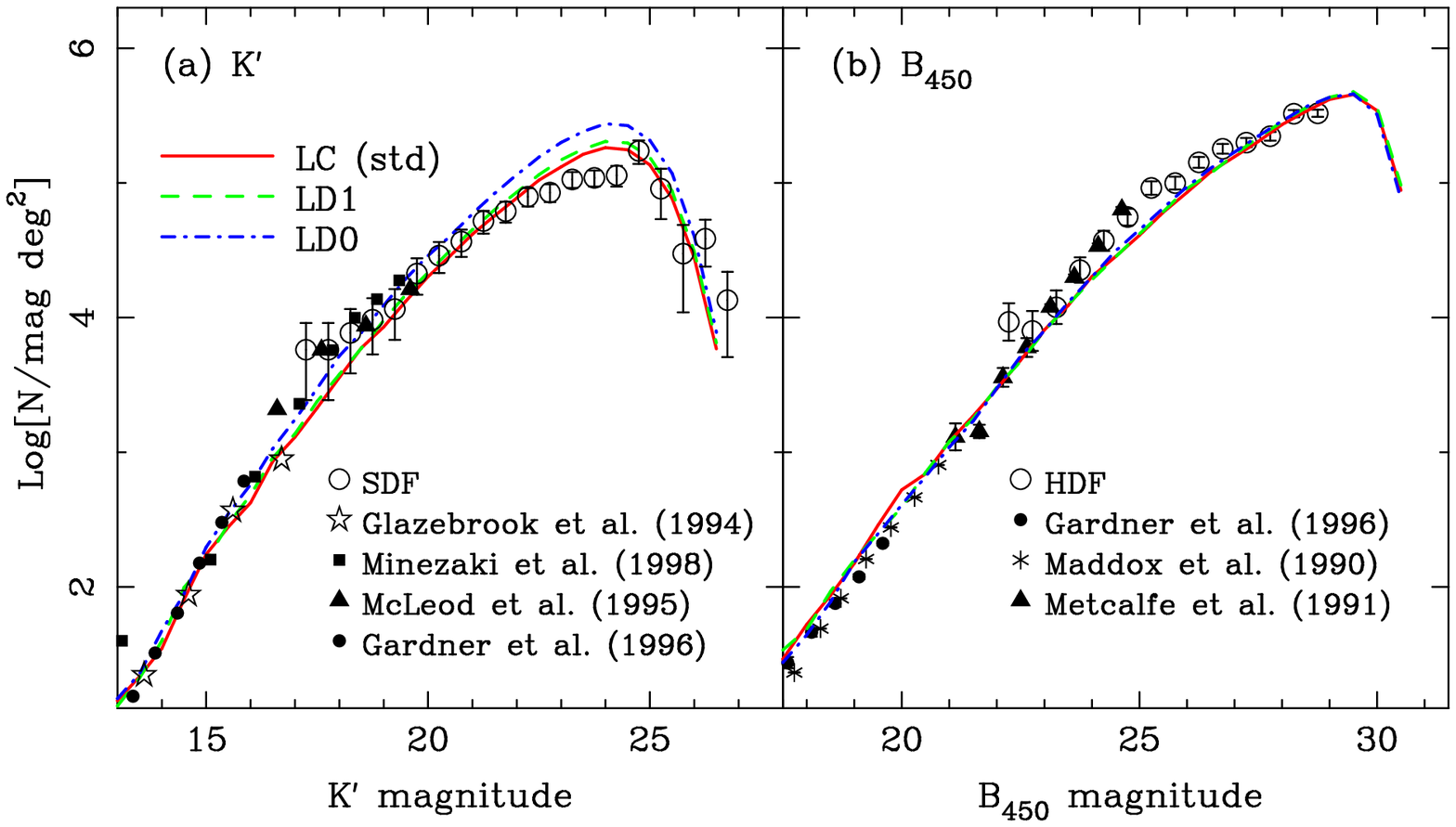}
\caption{Number-magnitude relations for a $\Lambda$-dominated flat CDM model.  
The solid, dashed, and dot-dashed lines indicate the LC (CSF), LD1 (DSF 
with $\sigma=1$), and LD0 (DSF with $\sigma=1$) models, respectively.  
}
\label{fig:sdf_sfr}
\end{figure}

Figure \ref{fig:sdf_sfr} shows the predicted number counts of galaxies
in the Subaru $K'$ and {\it HST} $B_{450}$ bands for the models of LC
(solid line), LD1 (dashed line) and LD0 (dot-dashed line). The predicted
counts from LD1 and LD0 are higher than those by LC; this tendency is
much more prominent in the longer wavelength passband.  This
wavelength-dependent tendency is explained as follows.  In general, the
near-infrared luminosity in the $K'$ band mainly reflects the total mass
of long-lived stars, while the UV/blue luminosity reflects the
instantaneous SFR.  While the total mass of cold gas available for star
formation is limited by the radiative cooling of hot gas, the conversion
timescale from cold gas to stars is determined by the SF timescale. In
the case of DSF, for which the SF timescale is shorter than in CSF, more
stars are formed at higher redshift.  Consequently, in the cases of LD1
and LD0, individual galaxies become brighter in the $K'$ band, and so
the $K'$-band counts of observable galaxies above the surface brightness
threshold become larger than in LC, although their blue luminosity is
hardly changed in the $B_{450}$ band.

The above argument indicates that the SF timescale is best constrained
by the observed SDF counts in the $K'$ band.  Since the LC model is
found to give a superior agreement with the SDF counts, the SF timescale
should therefore be nearly constant against redshift.  This result is
consistent with previous SAM analyses by \citet{kh00}, \citet{spf01} and
NTGY.

The optical depth for absorption of stellar light by dust is assumed to
be proportional to the amount of cold gas present.  In the cases of LD1
and LD0, because of the short SF timescale, the fraction of cold gas
mass at high redshift is smaller, leading to smaller optical depth when
compared to LC. For this reason, the $B_{450}$-band counts in LD1 and
LD0, which are sensitive to internal dust extinction, show a slight
excess at $B_{450}\sim 23$.  However this effect is almost canceled out
by the increase of stars due to the higher SFR given by the shorter SF
timescale, as shown in Figure \ref{fig:sdf_sfr}b.

\subsection{Galaxy Size}\label{sec:usize}

\begin{figure}
\plotone{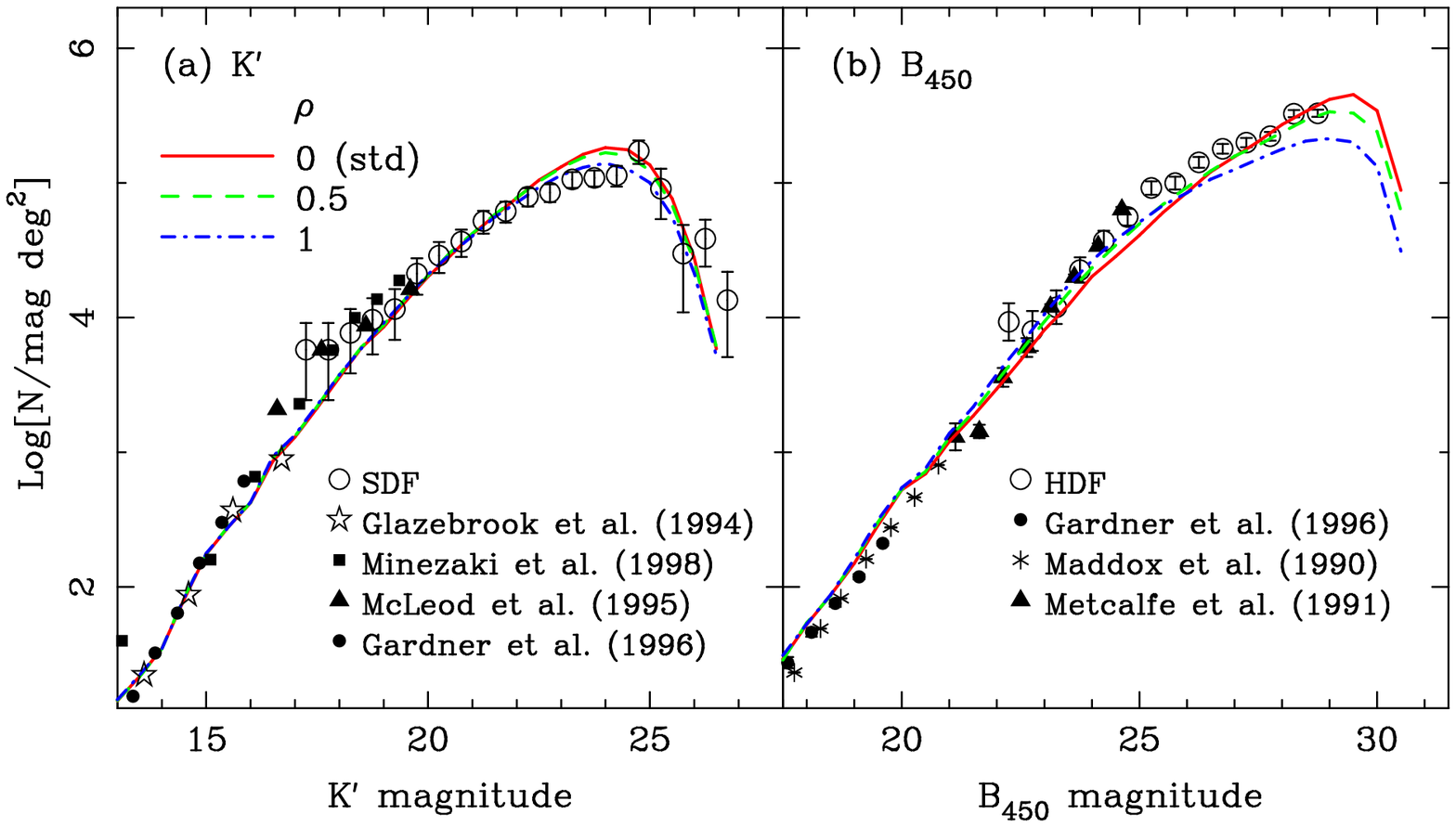}
\caption{Number-magnitude relations for a $\Lambda$-dominated flat CDM
model.  The solid, dashed, and dot-dashed lines indicate the LC models
with different $z$-dependences of galaxy size, $\rho=0$, 0.5, and
1, respectively.  }
\label{fig:sdf_rad}
\end{figure}

As mentioned in \S\ref{sec:counts}, the selection effects from the
cosmological dimming of surface brightness of galaxies are important.
Therefore, the number counts of galaxies are sensitive to their size,
because the surface brightness is proportional to the inverse of radius
squared.  Figure \ref{fig:sdf_rad} shows theoretical predictions for the
LC model with $\rho=0$ (solid line), 0.5 (dashed line), and 1
(dot-dashed line).  Note that the simple model with $\rho=0$ leads to
smaller disk size at high reshift in proportional to the initial radius
$r_{i}$, which is nearly equal to the virial radius in the case of halos
with galactic mass-scale and becomes smaller size according to the
spherical collapse model.  Thus the models with $\rho>0$ lead to larger
disks and lower surface brightness compared to the simple model with
$\rho=0$, and hence the smaller number counts of observable galaxies
above the surface brightness threshold, in both the $K'$ and $B_{450}$
bands, when compared to the standard case of $\rho=0$.  Since the case
of $\rho=0$ slightly overpredicts the $K'$-band observed counts at
$K'\sim 24$ and the case of $\rho\sim 0.5$ is preferable, it seems that
high-redshift galaxies might be more extended in size than expected from
the cooling radius.

The size of a galaxy disk also affects dust extinction. The optical
depth of dust is proportional to the inverse of disk radius squared.  In
the case of $\rho=1$, the number of galaxies at $B_{450}\sim 23$
increases above the $\rho=0$ prediction, because of weaker dust
extinction owing to larger disk size.  The effects of dust extinction
are discussed in the next section in detail.

The physical processes which cause the galaxy disks to be more extended at high
redshift are yet to be identified.  One possibility includes the viscous
evolution of the star-forming disk \citep{sy90}, but more investigations will
obviously be necessary here.

\subsection{Dust Extinction}\label{sec:udust}

The uncertainty in estimation of dust extinction at high redshift comes
from a combination of uncertainties in the chemical enrichment model,
dust-to-metal ratio, clumpiness of dust, and so on.  Instead of entering
into these details, however, we measure a robustness of our result by
changing the value of $\gamma$ in the optical depth by dust (equation
3).

\begin{figure}
\plotone{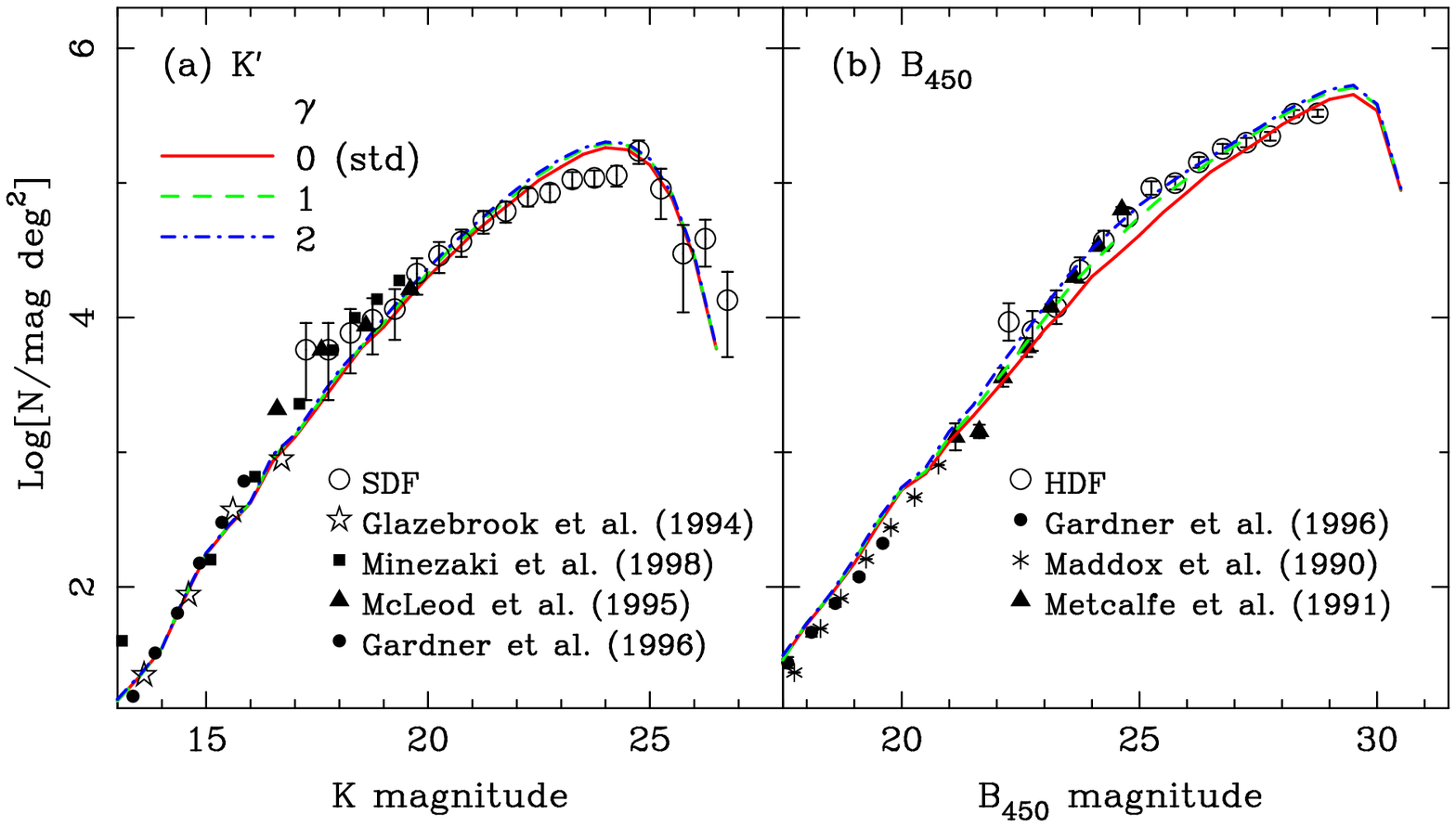} \caption{Number-magnitude relations for a
$\Lambda$-dominated flat CDM model.  The solid, dashed, and dot-dashed
lines indicate the LC models with different $z$-dependences of dust
optical depth, $\gamma=0$, 1, and 2, respectively, of the slab dust
model.  } \label{fig:sdf_opt}
\end{figure}

Figure \ref{fig:sdf_opt} shows theoretical predictions for the LC model
with $\gamma=0$ (solid line), 1 (dashed line), and 2 (dot-dashed line),
assuming the slab dust distribution .  The predicted $K'$-band counts
are almost independent of $\gamma$ (Figure \ref{fig:sdf_opt}a) because
the near-infrared passbands are insensitive to dust extinction.  On the
other hand, the predicted $B_{450}$-band counts {\it are} sensitive to
dust extinction (Figure \ref{fig:sdf_opt}b).  In the case of $\gamma\geq
1$, the ``knee'' in the count curve near $B_{450}\sim 24$ mag becomes
prominent.  It might be therefore necessary to adopt a high value of
$\gamma\ga 1$ in order to reproduce the observed $B_{450}$-band counts
better over a wide range of $B_{450}$-magnitude if the simple assumption
of disk size in the previous subsection is correct.

\begin{figure}
\plotone{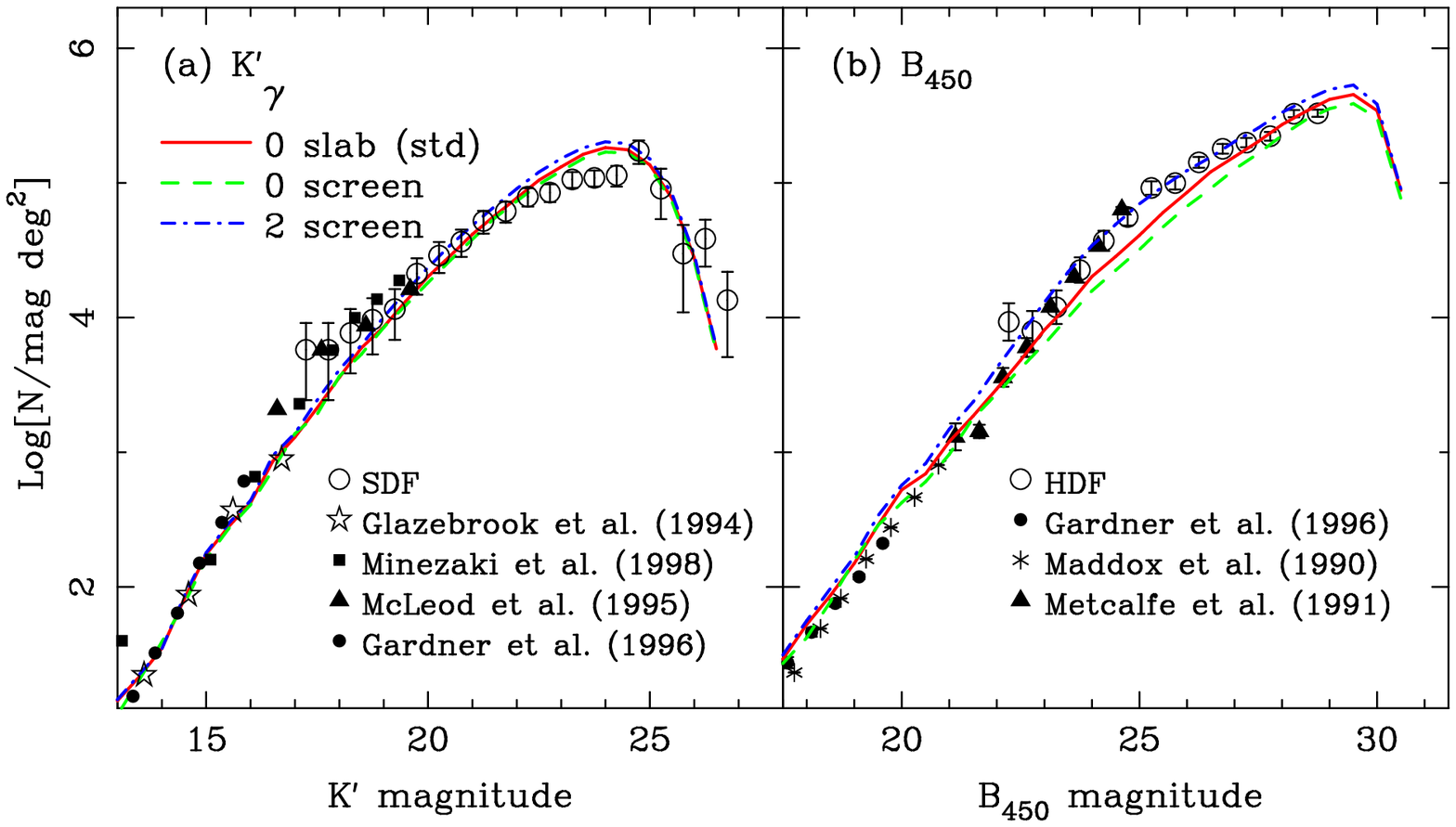}
\caption{Number-magnitude relations for a $\Lambda$-dominated flat CDM
model.  The solid, dashed, and dot-dashed lines indicate the LC models
with the {\it slab} dust with $\gamma=0$, and the {\it screen} dust with
$\gamma=0$ and 2, respectively.  }
\label{fig:sdf_slab}
\end{figure}

Figure \ref{fig:sdf_slab} is the same as Figure \ref{fig:sdf_opt},
except for showing theoretical predictions for the screen dust
distribution with $\gamma=0$ (dashed line) and 2 (dot-dashed line), in
comparison with the slab dust distribution with $\gamma=0$ (solid line).
In general, when the value of optical depth is fixed in the case of
$\gamma=0$, the screen dust model gives stronger extinction than the
slab dust model, so that the predicted number of bright galaxies at
$B_{450}\la 23$ becomes only slightly lower than that for the slab dust.
In the case of $\gamma=2$, the ``knee'' in the count curve near
$B_{450}\sim 24$ mag is prominent even in the screen dust model.  On the
other hand, the predicted $K'$-band counts are not affected at all, and
are therefore free from the uncertainty in estimation of dust
extinction.

The most effective discriminator of dust extinction is the knee of
observed galaxy counts at $B_{450}\sim 23$.  When the dust extinction is
strong, the knee disappears and the count curve deviates below the data.
Therefore, the optical depth by dust should be small at high redshift,
that is, $z\ga 1$.  We find that the knee is sensitive almost
exclusively to the optical depth, which is determined by the cold gas
mass, disk size, metallicity of cold gas, and $\gamma$ (equation
[\ref{eqn:dust}]).

In order to decrease the optical depth we need to decrease the cold gas
mass and/or metallicity, expand the galaxy size, or increase the value
of $\gamma$.  If one decreases the cold gas mass with a shorter SF
timescale, too many galaxies are formed, and the number of galaxies in
the $K'$-band is overpredicted (\S\ref{sec:usf}).  If one expands the
galaxy size, we suffer from too strong selection effects
(\S\ref{sec:usize}).  There is an uncertainty in estimation of the cold
gas metallicity \citep{spf01}, but this uncertainty can be practically
absorbed in the effect of $\gamma$.  Thus, in the usual SAMs
($\gamma=0$), a simple estimation of optical depth might be worse to
reproduce the observed galaxy counts at short wavelengths rather than
models with a high value of $\gamma\ga 1$.  In any case, the shape of
galaxy counts at shorter wavelength is sensitive to the optical depth.
We will need more investigations on the dust extinction.

\subsection{Morphology-Dependent Number Counts}\label{sec:morph}

Our SAM simultaneously reproduces the multi-band galaxy counts, in
contrast to the previous works using the traditional galaxy evolution
models \citep{ty00, t01}.  The reasons could indeed be quite
complicated, but one of the possibilities might be that a dominant
morphology of galaxies is different in different passband.  The SAM
allows for morphological transformations during the evolution of
galaxies, that is, mergers between galaxies of similar mass make
spheroidals as merger remnants, then they accrete hot gas through
radiative cooling as galactic disk.

Figure \ref{fig:sdf_morph} shows the predicted contributions of
early-type galaxies (E/S0; dashed lines) in the LC model in the $K'$ and
$I_{814}$ bands.  Note that the morphology of galaxies is identified by
the $B$-band bulge-to-disk ratio at their observed redshift in the
observer frame and that this definition is different from that in usual
analyses with the traditional models.  The total number counts are also
shown by the solid lines.  In order to see the fractional change of
morphologies, we change the value of $f_{\rm bulge}$ which divides
mergers into major and minor ones according to the mass ratio of merging
galaxies (see \S\ref{sec:gf}).  The thick lines are for the reference
value of $f_{\rm bulge}=0.1$, and the thin lines for $f_{\rm bulge}=0.5$
where major merger occurs only when the mass of smaller satellite is
larger than 0.5 times that of larger one.  We also plot the observed
E/S0 counts in the HDF \citep{a96, d98, p00} with the total HDF counts
in the $I_{814}$-band.  We find that our prediction for E/S0 counts
roughly agree with the observed data and that the uncertainty in the
E/S0 counts caused by that of $f_{\rm bulge}$ is nearly a factor of two.

We note that while the fraction of early-type galaxies increases with
decreasing $f_{\rm bulge}$, the total counts are hardly changed.  This
indicates that the total number of galaxies is essentially determined by
the merging histories of dark halos and by the cooling rate of hot gas,
and that the cooling must be very efficient to enable frequent formation
of disk galaxies at $z\ga 1$.

We also note that our morphological classification is based on the
$B$-band bulge-to-disk ratio.  It has not yet been confirmed that this
$B$-band classification coincides with that for the $K'$ band, therefore
our predictions of type-dependent counts in the $K'$ band should be regarded
as qualitative estimates.

\section{SUMMARY AND DISCUSSION}\label{sec:summary}

We have calculated the number counts of faint galaxies in the framework
of SAM for three cosmological models, the standard CDM (EdS) universe, a
low-density open universe, and a low-density flat universe with nonzero
$\Lambda$.  Our SAM includes the selection effects from the cosmological
dimming of surface brightness of galaxies with criteria appropriate for
the SDF and HDF, and also includes some modifications to our previous
analysis (NTGY), such as the optical depth estimation of dust within a
galaxy. In this paper we have shown that our SAM is fully consistent
with that of the previous version, and can explain the observed
multi-band galaxy counts from the UV to the near-infrared.

Comparison of theoretical predictions with the observed number counts of
SDF and HDF galaxies, as well as with other ground-based observations,
indicates that the standard CDM is ruled out, and a $\Lambda$-dominated
flat universe and a low-density open universe are favored. This result
is consistent with that from HDF galaxies (NTGY), but is in sharp
contrast with previous SAM analyses by other authors, where many of the
conceivable selection effects in the faint observations have been
ignored.

Some uncertainties in our SAM have been discussed.  These arise from a
lack of knowledge on the galaxy formation process, and also from an
insufficient survey of the physical properties of high-redshift
galaxies.  We especially focused on the uncertainties in redshift
dependence of SF timescale, galaxy size, and dust extinction.  We found
that dust extinction hardly affects galaxy counts in the Subaru $K'$
band, but does significantly affect those in the {\it HST} $B_{450}$
band.  Thus, the $K'$-band galaxy counts are robust against the
uncertainty of dust extinction.  Two other factors affect galaxy counts
even in the $K'$-band.  If the SF timescale at high redshift is shorter
than one which is simply proportional to the dynamical timescale in the
disk, too many galaxies are formed, and the number of galaxies at
faint-end is greatly overpredicted.  We found that the SF timescale
should be nearly constant against redshift, as suggested by our previous
analysis (NTGY) and by other recent SAM analyses\citep{kh00, spf01}.

The uncertainty in estimation of galaxy size results in an uncertainty
in estimation of the surface brightnesses of galaxies, which is directly
related to the selection effects mentioned above.  In our SAM, like
usual SAMs, the size of the disk is determined under the assumption of
specific angular momentum conservation of cooling gas, so that the disk
size is proportional to the cooling radius.  In order to see how
theoretical predictions are changed by changing the galaxy size, we
introduced a free parameter $\rho$ allowing for an additional redshift
dependence in size estimation.  We found that the value of $\rho\ga
0.5$ is favored in order to reproduce the observed counts, which indicates
that the disk radius should be extended by a factor of $(1+z)$ over the
cooling radius.  We also found that this manipulation cannot be
discriminated observationally because we cannot know how many undetected
low surface brightness galaxies below the selection criteria are there
at high redshift, which are presumably systems of large size.

Through this work, we have shown that our SAM can explain a variety of
observed properties of nearby and high-redshift galaxies, and place some
constraints on star formation, size evolution, and dust extinction.
More stringent constraints will certainly be obtained by a greater
knowledge of dynamical and kinematical properties of galaxies in near
future.

\acknowledgments

We would like to thank T. C. Beers for his critical reading of the
manuscript and T. Kodama for kindly giving us the data of the pass-bands
in the SDF.  We also thank the anonymous referee who led us to a
improvement of our paper.  This work has been supported in part by the
Grant-in-Aid for the Center-of-Excellence research (07CE2002) and for
the Scientific Research Funds (13640249 and 12047233) of the Ministry of
Education, Science, Sports and Culture of Japan.  Numerical computation
in this work was partly carried out at the Yukawa Institute Computer
Facility.

\clearpage

\begin{deluxetable}{cccccccccccc}
\tabletypesize{\scriptsize}
\tablecaption{Model Parameters\label{tab:astro}}  
\tablewidth{0pt}
\tablehead{
& \multicolumn{4}{c}{cosmological parameters} &
& \multicolumn{6}{c}{astrophysical parameters} \\
\cline{2-5} \cline{7-12}
\colhead{CDM Model} & \colhead{$\Omega_{0}$} &
 \colhead{$\Omega_{\Lambda}$} & \colhead{$h$} & \colhead{$\sigma_{8}$} & &
\colhead{$V_{\rm hot}$ (km~s$^{-1}$)}
& \colhead{$\alpha_{\rm hot}$} & \colhead{$\tau_{*}^{0}$ (Gyr)}
& \colhead{$\alpha_*$} & \colhead{$f_{\rm b}$} 
& \colhead{$\Upsilon$}}
\startdata
SC & 1 & 0 &0.5&0.67&& 350 & 5   & 7   & -3   & 2 & 1\\
OC &0.3& 0 &0.6& 1  && 240 & 3   & 1.8 & -2.2 & 1 & 2\\
LC &0.3&0.7&0.7& 1  && 280 & 2.5 & 2   & -2   & 1 & 1.5\\
\enddata
\tablecomments{Equations defined the SN feedback-related parameters
 ($V_{\rm hot}$ and $\alpha_{\rm hot}$) and the SFR-related parameters
 ($\tau_{*}^{0}$ and $\alpha_{*}$) are (2) and (1), respectively.  The
 definitions of last two parameters, $f_{\rm b}$ and $\Upsilon$, are
 given in \S\S 2.3 and 2.2, respectively.  The other parameters used in
 common as standard include $\Omega_{\rm b}=0.015h^{-2}$, CSF (equation
 1), $f_{\rm bulge}=0.1$ (\S2.1), $\gamma=0$ (equation 3), $\rho=0$
 (equation 6), and ($\bar{\lambda}, \sigma_{\lambda}$)=(0.05, 0.5)
 (equation 5).}
\end{deluxetable}


\begin{thebibliography}{}   
\bibitem[Abraham et al.(1996)]{a96}Abraham, R.G., Tanvir, N.R.,
			      Santiago, B.X., Ellis, R.S., Glazebrook,
			      K., \& van den Bergh, S. 1996, \mnras,
			      279, 47
\bibitem[Arimoto \& Yoshii(1986)]{ay86}Arimoto, N., \& Yoshii, Y. 1986,
                                 \aap, 164, 260
\bibitem[Arimoto \& Yoshii(1987)]{ay87}Arimoto, N., \& Yoshii, Y. 1987,
                                 \aap, 173, 23
\bibitem[Arimoto, Yoshii \& Takahara(1991)]{ayt91}Arimoto, N., Yoshii,
                                 Y., \& Takahara, F. 1991, \aap, 253, 21
\bibitem[Baugh, Cole \& Frenk(1996)]{bcf96}Baugh, C. M., Cole, S., \&
                                  Frenk, C. S. 1996, \mnras, 283, 1361
\bibitem[Baugh et al.(1998)]{b98}Baugh, C. M., Cole, S., Frenk, C. S.,
                                  \& Lacey, C. G. 1998, \apj, 498, 504
\bibitem[Binney \& Tremaine(1987)]{bt87}Binney, J., \& Tremaine,
                                 S. 1987, Galactic Dynamics, Princeton
                                 Univ. Press, Princeton, NJ
\bibitem[Blanton et al.(2001)]{b01}Blanton, M.R. et al. 2001, \aj, 121, 2358
\bibitem[Bond et al.(1991)]{bcek91}Bond, J. R., Cole, S., Efstathiou,
                                 G., \& Kaiser, N. 1991, \apj, 379, 440
\bibitem[Bower(1991)]{b91}Bower, R. 1991, \mnras, 248, 332
\bibitem[Catelan \& Theuns(1996a)]{ct96a}Catelan, P., \& Theuns, T. 1996a,
                                 \mnras, 282, 436
\bibitem[Catelan \& Theuns(1996b)]{ct96b}Catelan, P., \& Theuns, T. 1996b,
                                 \mnras, 282, 455
\bibitem[Cole et al.(1994)]{c94}Cole, S., Aragon-Salamanca, A., Frenk,
                                  C. S., Navarro, J. F., \& Zepf,
                                  S. E. 1994, \mnras, 271, 781
\bibitem[Cole et al.(2000)]{c00}Cole, S., Lacey, C. G., Baugh, C. M., \&
                                  Frenk, C. S. 2000, \mnras, 319, 168
\bibitem[Cole et al.(2001)]{c01}Cole, S. et al. 2001, \mnras, 326, 255
\bibitem[Driver et al.(1998)]{d98}Driver, S.P., Fern{\'a}ndez-Soto, A.,
				 Couch, W.J., Odewahn, S.C., Windhorst,
				 R.A., Phillipps, S., Lanzetta, K., \&
				 Yahil, A. 1998, \apj, 496, 93
\bibitem[Eke, Cole \& Frenk(1996)]{ecf96}Eke, V. R., Cole, S., \& Frenk,
                                 C. S. 1996, \mnras, 282, 263
\bibitem[Fall(1979)]{f79}Fall, S. M. 1979, \nat, 281, 200
\bibitem[Fall \& Efstathiou(1980)]{fe80}Fall, S. M., \& Efstathiou,
			      G. 1980, \mnras, 193, 189
\bibitem[Fall(1983)]{f83}Fall, S. M. 1983, in `Internal kinematics and
                                 dynamics of galaxies', proceedings of
                                 the IAU symposium 100, Besancon,
                                 France, Dordrecht, D. Reidel, p.391
\bibitem[Fern{\'a}ndez-Soto, Lanzetta \&
                                 Yahil(1999)]{fly99}Fern{\'a}ndez-Soto,
                                 A., Lanzetta, K. M., \& Yahil, A. 1999,
                                 \apj, 513, 34
\bibitem[Folkes et al.(1999)]{f99}Folkes, S. et al. 1999, \mnras, 308, 459
\bibitem[Furusawa et al.(2000)]{f00}Furusawa, H., Shimasaku, K., Doi,
                                 M., \& Okamura, S. 2000, \apj, 534, 624
\bibitem[Gardner et al.(1996)]{gscf96}Gardner, J. P., Sharples, R. M.,
                                 Carrasco, B. E., \& Frenk, C. S. 1996,
                                 \mnras, 282, L1
\bibitem[Gardner et al.(1997)]{gsfc97}Gardner, J. P., Sharples, R. M.,
                                 Frenk, C. S., \& Carrasco, B. E. 1997,
                                 \apjl, 480, L99
\bibitem[Glazebrook et al.(1994)]{g94}Glazebrook, K., Peacock, J.A.,
                                 Miller, L., \& Collins, C.A. 1994,
                                 \mnras, 266, 65
\bibitem[Gunn \& Gott(1972)]{gg72}Gunn, J.E., \& Gott, J.R. 1972, ApJ,
                                 176, 1
\bibitem[Hall \& Mackay(1984)]{hm84}Hall, P., \& Mackay, C. B. 1984,
                                 \mnras, 210, 979
\bibitem[Heyl et al.(1995)]{h95}Heyl, J. S., Cole, S., Frenk, C. S., \&
                                  Navarro, J. F. 1995, \mnras, 274, 755
\bibitem[Huchtmeier \& Richter(1988)]{hr88}Huchtmeier, W. K., \&
                                 Richter, O. -G. 1988, \aap, 203, 237
\bibitem[Impey et al.(1996)]{isib96}Impey, C.D., Sprayberry, D., Irwin,
                                 M. J., \& Bothun, G. D.        1996,
                                 \apjs, 105, 209
\bibitem[Jones et al.(1991)]{jfsep91}Jones, L.R., Fong, R., Shanks, T.,
                                 Ellis, R. S., \& Peterson, B. A. 1991,
                                 \mnras, 249, 481
\bibitem[Kauffmann \& Haehnelt(2000)]{kh00}Kauffmann, G., \& Haehnelt,
                                 M. 2000, \mnras, 311, 576
\bibitem[Kauffmann, White \& Guiderdoni(1993)]{kwg93}Kauffmann, G.,
                                  White, S. D. M., \& Guiderdoni,
                                  B. 1993, \mnras, 264, 201
\bibitem[Kauffmann, Guiderdoni \& White(1994)]{kgw94}Kauffmann, G.,
                                 Guiderdoni, B., \& White,
                                 S. D. M. 1993, \mnras, 267, 981
\bibitem[Kodama \& Arimoto(1997)]{ka97}Kodama, T., \& Arimoto, N. 1997,
                                  \aap, 320, 41
\bibitem[Koo(1986)]{k86}Koo, D.C. 1986, \apj, 311, 651
\bibitem[Lacey \& Cole(1993)]{lc93}Lacey, C.G., \& Cole, S. 1993,
                                 \mnras, 262, 627   
\bibitem[Lanzetta et al.(2001)]{lypcf}Lanzetta, K.M., Yahata, N.,
			      Pascarelle, S., Chen, H.-W., \&
			      Fern\'{a}ndez-Soto, A. 2002, preprint,
			      astro-ph/0111129
\bibitem[Loveday et al.(1992)]{l92}Loveday, J., Peterson, B. A.,
                                 Efstathiou, G., \& Maddox, S. J. 1992,
                                 \apj, 90, 338
\bibitem[Maddox et al.(1990)]{mselp90}Maddox, S. J., Sutherland, W. J.,
                                 Efstathiou, G., Loveday, J., \&
                                 Peterson, B. A. 1990, \mnras, 247, 1p
\bibitem[Maihara et al.(2001)]{m01}Maihara, T. et al. 2001, \pasj, 53,
                                 25
\bibitem[Makino \& Hut(1997)]{mh97}Makino, J., \& Hut, P. 1997, \apj,
                                 481, 83
\bibitem[McLeod et al.(1995)]{mbrtf95}McLeod, B.A., Bernstein, G.M.,
                                 Rieke, M.J., Tollestrup, E.V., \&
                                 Fazio, G.G. 1995, \apjs,  96, 117
\bibitem[Metcalfe et al.(1991)]{msfj91}Metcalfe, N., Shanks, T., Fong,
                                 R., \& Jones, L. R. 1991, \mnras, 249, 498
\bibitem[Minezaki et al.(1998)]{mkyp98}Minezaki, T., Kobayashi, Y.,
                                 Yoshii, Y., \& Peterson, B.A. 1998,
                                 \apj, 494, 111
\bibitem[Mo, Mao \& White(1998)]{mmw98}Mo, H.J., Mao, S., \& White,
                                 S.D.M. 1998, \mnras, 295, 319
\bibitem[Mobasher, Sharples \& Ellis(1993)]{mse93}Mobasher, B.,
                                 Sharples, R. M., \& Ellis, R. S. 1993,
                                 \mnras, 263, 560
\bibitem[Nagashima \& Gouda(1997)]{ng97}Nagashima, M., \& Gouda,
                                 N. 1997, \mnras, 301, 849
\bibitem[Nagashima et al.(2001)]{n01}Nagashima, M., Totani, T., Gouda,
                                 N., \& Yoshii, Y. 2001, \apj, 557, 505
                                 (NTGY)
\bibitem[Netterfield et al.(2001)]{boomerang}Netterfield, C. B. et
			      al. 2001, preprint, astro-ph/0104460
\bibitem[Peebles(1993)]{p93}Peebles, P. J. E. 1993, Principles of
                                 Physical Cosmology, Princeton
                                 Univ. Press, Princeton, NJ
\bibitem[Phillipps et al.(2000)]{p00}Phillipps, S., Driver, S.P., Couch,
				 W.J., Fern{\'a}ndez-Soto, A., Bristow,
				 P.D., Odewahn, S.C., Windhorst, R.A.,
				 \& Lanzetta, K. 2000, \mnras, 319, 807
\bibitem[Press \& Schechter(1974)]{ps74}Press, W., \& Schechter,
                                 P. 1974, \apj, 187, 425
\bibitem[Ratcliffe et al.(1998)]{r98}Ratcliffe, A., Shanks, T., Parker,
                                 Q., \& Fong, R. 1998, \mnras, 293, 197
\bibitem[Saio \& Yoshii(1990)]{sy90}Saio, H., \& Yoshii, Y. 1990, \apj,
                                 363, 40
\bibitem[Simien \& de Vaucouleurs(1986)]{sdv86}Simien, F., \& de
                                 Vaucouleurs, G. 1986, \apj, 302, 564
\bibitem[Somerville \& Kolatt(1999)]{sk99}Somerville, R.S., \& Kolatt,
                                  T. 1999, \mnras, 305, 1
\bibitem[Somerville \& Primack(1999)]{sp99}Somerville, R.S., \&
                                  Primack, J. R. 1999, \mnras, 310, 1087
\bibitem[Somerville, Primack \& Faber(2001)]{spf01}Somerville, R.S.,
                                 Primack, J. R., \& Faber, S. M. 2001,
                                 \mnras, 320, 504
\bibitem[Sutherland \& Dopita(1993)]{sd93}Sutherland, R., \& Dopita,
                                  M. A. 1993, \apjs, 88, 253
\bibitem[Suzuki, Yoshii \& Beers(2000)]{syb00}Suzuki, T.K., Yoshii,
                                  Y., \& Beers, T.C. 2000, \apj, 540, 99
\bibitem[Tomita(1969)]{t69}Tomita, K. 1969, Prog. Theor. Phys., 42, 9
\bibitem[Totani \& Yoshii(2000)]{ty00}Totani, T., \& Yoshii, Y. 2000,
                                  \apj, 540, 81
\bibitem[Totani et al.(2001a)]{t01}Totani, T., Yoshii, Y., Maihara, T.,
                                 Iwamuro, F., \& Motohara, K. 2001a,
                                 \apj, 559, 592
\bibitem[Totani et al.(2001b)]{tyimm}Totani, T., Yoshii, Y., Iwamuro, F.,
			      Maihara, T., \& Motohara, K. 2001b, \apjl,
			      558, L87
\bibitem[Totani \& Takeuchi(2002)]{tt}Totani, T., \& Takeuchi,
			      T.T. 2002, \apj in press (astro-ph/0201277)
\bibitem[Tyson(1988)]{t88}Tyson, J.A. 1988, \aj, 96, 1
\bibitem[White(1984)]{w84}White, S.D.M. 1984, \apj, 286, 38
\bibitem[Williams et al.(1996)]{w96}Williams, R. T. et al. 1996, AJ,
                                  112, 1335
\bibitem[Yoshii(1993)]{y93}Yoshii, Y. 1993, \apj, 403, 552
\bibitem[Yoshii \& Arimoto(1987)]{ya87}Yoshii, Y., \& Arimoto, N. 1987,
                                 \aap, 188, 13
\bibitem[Yoshii \& Peterson(1991)]{yp91}Yoshii, Y., \& Peterson,
                                  B. A. 1991, \apj, 372, 8
\bibitem[Yoshii \& Peterson(1994)]{yp94}Yoshii, Y., \& Peterson,
                                  B. A. 1994, \apj, 436, 551
\bibitem[Yoshii \& Peterson(1995)]{yp95}Yoshii, Y., \& Peterson,
                                  B. A. 1995, \apj, 444, 15
\bibitem[Yoshii \& Takahara(1988)]{yt88}Yoshii, Y., \& Takahara,
                                  F. 1988, \apj, 326, 1
\bibitem[Zucca et al.(1997)]{z97}Zucca, E. et al. 1997, \aap, 326, 477
\end{thebibliography}
\end{document}